# Moisture-driven CO$_2$ direct air capture and delivery for cultivating cyanobacteria

Justin Flory,[a] Shuqin Li,[b] Samantha Taylor,[a] Sunil Tiwari,[a] Garrett Cole,[d] Marlene Velazco Medel,[a] Amory Lowe,[a] Jordan Monroe,[c] Sara Sarbaz,[e] Nick Lowery,[a] Joel Eliston,[a] Heidi P. Feigenbaum,[e] Heather Emady,[c] Jason C. Quinn,[d] Matthew Green,[a,c] John McGowen,[f] Klaus Lackner,[a] Wim Vermaas[b]

[a]*Center for Negative Carbon Emissions at Arizona State University (ASU), 777 E. University Dr., Tempe, Arizona, 85281, USA*
[b]*School of Life Sciences at ASU, Tempe, Arizona, 85287-4501, USA*
[c]*School for Engineering of Matter, Transport and Energy at ASU, Tempe, Arizona, 85287-4501, USA*
[d]*Sustainability Sciences LLC, Steamboat Springs, Colorado 80487, USA*
[e]*Department of Mechanical Engineering, Northern Arizona University, Flagstaff, Arizona, 86011, USA*
[f]*Center for Algae Technology and Innovation (AzCATI) at ASU, 7418 E. Innovation Way South, Mesa, Arizona, 85234, USA*

## 1. Abstract

A moisture-driven air capture (DAC) system was developed and demonstrated for cultivating cyanobacteria and microalgae at the flask (50 mL), bench (12 L) and small pilot (840 L) scale. Purolite A501 anion exchange resin beads, enclosed in mesh packets, were found to be biocompatible and rapidly deliver air-captured CO$_2$ when immersed directly in an alkaline cultivation medium containing cyanobacteria or microalgae. Flask-scale cultivation trials showed A501 could sustain rapid growth (190 mg L$^{-1}$ d$^{-1}$) of the cyanobacterium *Synechocystis* sp. PCC 6803 strain engineered to produce laurate (*Synechocystis* TE/Δ*slr1609*). A bench-scale system installed in a laminar flow hood was able to deliver ~ 2 g CO$_2$ d$^{-1}$ into abiotic alkaline cultivation medium and ~0.5 g d$^{-1}$ in the presence of *Synechocystis* TE/Δ*slr1609* to support vigorous growth (39 mg L$^{-1}$ d$^{-1}$) limited by the CO$_2$ delivered by the sorbent. A small pilot-scale system installed in a 4.2 m$^2$ outdoor raceway pond in Mesa, Arizona was able to deliver ~100 g CO$_2$ d$^{-1}$ into abiotic alkaline cultivation medium. During outdoor cultivation with wild type *Synechocystis* sp. PCC 6803 (*Synechocystis* 6803), exopolysaccharides and other products excreted by *Synechocystis* 6803 covered the sorbent beads, reducing their capacity to ~25%, which could be partially restored to ~70% capacity using a wash protocol of 0.1% peracetic acid followed by 0.04% (w/v) bleach, 1 mM HCl and ion exchange in NaHCO$_3$, but the CO$_2$ delivery kinetics remained 3-4 fold slower. Analysis of the sorbent beads used as part of four separate outdoor cultivation trials with over 300 days of outdoor wet/dry cycling over four seasons showed significant mechanical fracturing; we hypothesize this fracturing was due to mechanical stresses imparted by water gradients during wet/dry cycles, exacerbated by reduced water transport from biofouling clogging the bead pores. Infrared spectroscopy and thermogravimetric analysis showed a significant loss of NR$_4^+$ functional groups necessary for CO$_2$ capture correlated with extended use. Under the assumption that abiotic performance eventually can be retained by delivering CO$_2$ into the media recycle stream in a way that avoids biofouling, technoeconomic and life cycle analyses show the viability of a small first-of-a-kind biorefinery producing 500 barrels per day of biofuel. The products include extracting the blue pigment-protein complex phycocyanin valued at $50 kg$^{-1}$ as natural food and beverage dye, extracting the remaining protein valued at $6 kg$^{-1}$ as a protein supplement and the remaining biomass is hydrothermally treated into biofuel valued at $2.50 per gasoline gallon equivalent.

**Terms and abbreviations**

| | |
|---|---|
| AEM | Anion Exchange Membrane |
| AER | Anion Exchange Resin |
| *1g* | Moisture-driven direct air capture system designed to delivery ~1 g $CO_2$ per day |
| *100g* | Moisture-driven direct air capture system designed to delivery ~100 g $CO_2$ per day |
| AzCATI | Arizona Center for Algae Technology and Innovation |
| DI | Deionized |
| DAC | Direct Air Capture |
| IC | Inorganic carbon |
| $OD_{730}$ | Optical density at 730 nm |
| RH | Relative humidity |
| *Synechocystis* 6803 | *Synechocystis* sp. PCC 6803 – wild type |
| *Synechocystis* TE/Δ*slr1609* | *Synechocystis* sp. PCC 6803 – strain engineered to produce laurate |
| QA | Quaternary Ammonium ($NR_4^+$) |

**Highlights**

- Biocompatible moisture-driven sorbents delivered $CO_2$ captured from ambient air to support vigorous growth of cyanobacteria at the bench scale
- Anion exchange resin beads showed significant mechanical fracturing and loss of $NR_4^+$ groups responsible for $CO_2$ capture after being exposed to four outdoor cultivation trials with cyanobacteria and microalgae with over 300 days of outdoor use and wet/dry cycling over four seasons in Mesa, Arizona.

**2. Introduction**

Photosynthetic organisms, including plants, algae and cyanobacteria, capture carbon dioxide ($CO_2$) from atmospheric air to produce a diverse array of biomass that can be converted into biofuels, sustainable chemicals and other materials that can be used in lieu of fossil-based products to mitigate rising atmospheric $CO_2$ levels.[1] However, plants require significant arable land, water and nutrients that limit the net climate benefit of bio-based products.[1] Microalgae and cyanobacteria can be cultivated using non-potable water (e.g., wastewater, brackish water, saltwater) and on non-arable land to create biomass much faster than plants and can be engineered to directly produce valuable chemicals and fuels with higher selling prices.[2] However, concentrated $CO_2$ sources are needed to maximize productivity since atmospheric levels of $CO_2$ limit carbon fixation and production rates of biomass, biofuel, and bioproducts, particularly for aquatic organisms.[3] Traditionally, to cultivate photosynthetic microbes, concentrated fossil-based $CO_2$ is delivered by sparging; however, in shallow raceway ponds >70% of the sparged $CO_2$ is released into the atmosphere, which significantly increases the cost of $CO_2$ utilized by the microbes and significantly reduces the net climate benefit.[4] Further, sunny locations ideal for cultivating photosynthetic microbes are often not near concentrated $CO_2$ sources, requiring costly and socially contentious pipelines or even more costly $CO_2$ transport by truck or rail. Cultivating at high pH and alkalinity can enable $CO_2$ capture directly from air into the cultivation medium, but this approach is limited to cyanobacteria and alkaliphilic microalgae and the bioavailability of $CO_2$ at high pH can limit growth.[5–7]

Direct air capture (DAC) technologies can produce concentrated $CO_2$ streams at the location where $CO_2$ is needed. DAC technologies largely use liquid or solid sorbent materials with strong binding affinities for $CO_2$ to capture $CO_2$ directly from atmospheric air to produce concentrated $CO_2$ streams suitable for sequestration in geologic formations, minerals and building materials, conversion into sustainable fuels or chemicals, or enhancing growth of plants (in greenhouses), microalgae or cyanobacteria.[8] However, most DAC systems use energy-intensive fans to move dilute air streams and heat or vacuum pressure to overcome the strong binding affinity of $CO_2$ to the sorbent material, with energy use far exceeding the thermodynamic requirements.[9–15] Lackner and Wright showed that strong-base anion exchange resins (AER) containing fixed cations (e.g., quaternary ammonium; $NR_4^+$) exchanged with $HCO_3^-$, $CO_3^{2-}$ or $OH^-$ anions capture $CO_2$ from ambient air when dry. When the AER is then exposed to water vapor or liquid water, $CO_2$ is released to increase the partial pressure of $CO_2$ up to 500-fold through a moisture-driven process.[16,17] $CO_3^{2-}$ remaining on the resin can then restart the capture cycle (see Figure 1 of Flory et al).[18]

Here, we describe a process for cultivating photosynthetic microbes using moisture-driven DAC sorbents that are continuously cycled between air to capture $CO_2$ and a liquid alkaline growth medium where $CO_2$ is released and stored until it is consumed during photosynthesis (**Figure 1**). First, a series of AER sorbents were evaluated for biocompatibility, $CO_2$ capacity and delivery performance. The best performing sorbent A501 was used to cultivate a strain of the cyanobacterium *Synechocystis* sp. PCC 6803 engineered in the Vermaas lab to produce laurate as a fuel and chemical feedstock[19] (*Synechocystis* TE/Δ*slr1609*) in 50 mL flasks followed by a bench-scale DAC system capable of delivering ~1 g $CO_2$ per day (called *1g*). Next a larger prototype DAC system capable of delivering ~100 g $CO_2$ per day (called *100g*) was demonstrated for cultivating a wild-type *Synechocystis* sp. PCC 6803 (*Synechocystis* 6803) in 4.2 m² outdoor raceway ponds. A protocol was developed to remove biofouling on the sorbent and mesh packet to partially restore its $CO_2$ capture and delivery performance. Chemical and physical changes to the sorbent were analyzed to determine the causes of fracturing and permanent $CO_2$ performance losses after up to 300 days of use in outdoor trials. Technoeconomic and life cycle analyses (TEA/LCA) evaluated the feasibility of a small first-of-a-kind biorefinery.

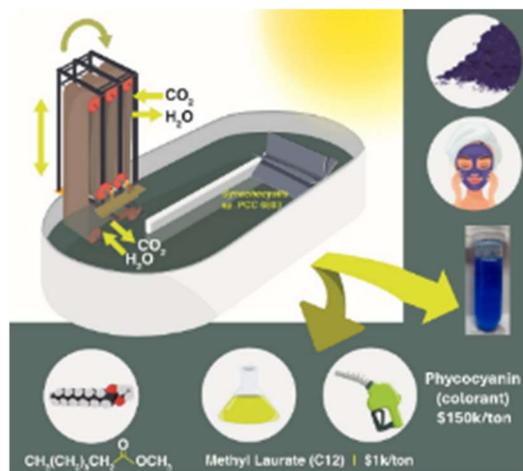

**Figure 1.** Illustration of the *100g* concept using moisture-driven $CO_2$ sorbents to deliver $CO_2$ from the air to liquid medium for cultivating cyanobacteria to produce sustainable fuels, chemicals and high-value products.

## 3. Materials and Methods

### 3.1 Sorbent Materials

Sorbent preparation and ion exchange: AERs with ion exchange capacities of ~2 meq per gram (dry basis) were purchased in the chloride form, which is not active for moisture-driven DAC, including beaded sorbents A501 (Purolite), HPR-4800 (DuPont), and Amberlite IRA-900 (DuPont), and one sheet sorbent Excellion (SnoPure). The desired quantity of sorbent was added to a fresh solution of 1.0 M $NaHCO_3$ (baking soda) and soaked overnight while gently stirring this process was repeated once, and then the sorbent was rinsed multiple times in deionized (DI) water to remove excess $NaHCO_3$. Lower concentrations of $NaHCO_3$ are also suitable as long as they have at least a 20−30-fold molar excess over the number of charge sites in the AER to be exchanged (e.g., 1 g sorbent in ≥ 100 mL of 0.5 M $NaHCO_3$).[18]

Measuring ion exchange capacity: The ion exchange capacities of several commercial AERs, including A501 (Purolite), HPR-4800 (DuPont), and Amberlite IRA-900 (DuPont), were measured using Mohr's method.[20] Briefly, 2 g of each resin in the chloride form were added to separate 500 mL solutions of 0.1 M $NaHCO_3$ to which 20 mL of 0.1 M potassium chromate ($K_2CrO_4$) had been added as the indicator. Small aliquots of a 0.1 M solution of silver nitrate ($AgNO_3$) was added to the $NaHCO_3$ + $K_2CrO_4$ solution until the solution's color changed from yellow to orangish-red. The amount of $AgNO_3$ added up to when the color changed was used to determine the amount of chloride released by the resin and thus its ion exchange capacity.

Abiotic test: Sorbent packet $CO_2$ binding capacity and delivery kinetic evaluation: Mesh packets with 1 g of sorbent beads were dried overnight in ambient air, submerged in 1 L of 10 mM $Na_2CO_3$, and the pH of the solution was recorded every 30 seconds for ~1 hour with constant stirring using a magnetic stirrer at 100 rpm. The pH change was converted to the amount of $CO_2$ delivered by the sorbent using calculations described by Shesh et al.[21]

## 3.2 Bench-scale cultivation and analysis

Standard BG-11 medium: Standard BG-11 medium used for initial cultivation trials and bringing the cultures up from plate includes 17.6 mM $NaNO_3$, 0.175 mM $K_2HPO_4$, 0.30 mM $MgSO_4$, 0.24 mM $CaCl_2$, 0.03 mM citric acid, 0.023 mM ferric ammonium citrate, 0.0027 mM disodium ethylenediaminetetraacetic acid (EDTA), 0.19 mM $Na_2CO_3$, and trace minerals.

Sorbent biocompatibility and cultivation: A mutant strain of *Synechocystis* 6803 modified to secrete laurate (*Synechocystis* TE/Δ*slr1609*)[19] was inoculated from colonies on a plate in standard BG-11 medium supplemented with 20 mM $NaHCO_3$. One g of each beaded sorbent in the mesh bag was immersed in a 50-mL culture of the *Synechocystis* TE/Δ*slr1609* strain grown on a shaker at 125 rpm at 30 °C under LED lights at a light intensity of 150 µmol photons $m^{-2}\,s^{-1}$; an additional 20 mM $NaHCO_3$ was added daily. The sorbents were retained in the cultures during the entire growth period. After the study, the sorbents were cleaned with 0.04% (w/v) sodium hypochlorite and then exchanged with 0.5 M $NaHCO_3$ to regenerate them for future use.

Optical density and $CO_2$ biomass capture analysis: One-mL aliquots of culture were sampled every 12–24 hours to monitor growth by measuring the optical density at 730 nm ($OD_{730}$) using a Shimadzu UV-1800 spectrophotometer. The amount of $CO_2$ fixed by *Synechocystis* cells was calculated based on the accumulated biomass. For *Synechocystis*, an $OD_{730}$ of 1.0 corresponds to 0.20–0.25 g/L dry biomass. Since carbon makes up ~50% of the dry cell weight, using the higher value of 0.25 g/L, a 12-L culture with an $OD_{730}$ of 1.0 fixed 5.5 g $CO_2$.

Modified BG-11 medium: A modified BG-11 medium was prepared for cultivation with lower levels of $NaNO_3$ (8.8 mM for bench scale and 5 mM for open pond) to minimize ion exchange with the resin; standard nitrate levels (17.6 mM) are in significant excess of what is needed for cell growth. The other BG-11 components were standard. To this modified BG-11 medium, 35 mM $NaHCO_3$ was added to provide a buffer to store captured $CO_2$ and maintain an excess of moisture-driven DAC active anions ($HCO_3^-$, $CO_3^{2-}$, $OH^-$) to inactive anions in the medium ($NO_3^-$, $Cl^-$, $SO_4^{2-}$, and $HPO_4^{2-}$) to ensure that ~80% of the anions in solution (and in equilibrium with the AER) are moisture-swing active.

Moisture-driven DAC flask scale cultivation: One g A501 (Purolite) or HPR-4800 (DuPont) sorbent was placed in a mesh bag with 25 µm diameter pores, immersed in BG-11 medium for 30 minutes to allow the sorbent to release $CO_2$ into the medium, and then, after removing the mesh bag, *Synechocystis* TE/Δ*slr1609* cells were added to the growth medium. The sorbent in the mesh bag was rinsed with deionized (DI) water thoroughly and then left dry in the laminar flow hood overnight to load $CO_2$ for subsequent use. The sorbent in the mesh bag was soaked each day for 30 minutes in a 50 mL culture of *Synechocystis* TE/Δ*slr1609* in standard or modified BG-11 medium to release $CO_2$ into the medium.

*1g* system design and installation: The *1g* system was designed and constructed as described by Flory et al.[18] out of 80/20 T-slot framing to support a motor-driven chain circulating a sorbent belt installed in a laminar flow hood, with an LED light panel (Samsung; SI-B8R341B20WW) placed underneath the 12-L capacity basin to provide light (50 or 250 µmol photons $m^{-2}\,s^{-1}$) to grow cyanobacteria without shading from the *1g* system; two 100 W aquarium heaters (Freesea) were placed in the basin to maintain the temperature of the growth medium at 30 °C. As described in Flory et al.,[18] the belt was made of 16 nylon mesh packets (LBA), each filled with 1 g of A501 (Purolite) AER beads. The belt was slowly rotated to expose the packets to the air for capturing $CO_2$ and to the alkaline modified BG-11 medium to release and retain the captured $CO_2$ as $HCO_3^-$ until it was consumed by cyanobacteria.

Inorganic carbon analysis: 50 mL aliquots of the cultivation medium were sampled daily in a sealed tube. Then 2 N HCl was added 1:100 to the sample to drive off all inorganic carbon as $CO_2$, which was measured using non-dispersive infrared detection in a Shimadzu TOC/TN analyzer.

## 3.3 Pond-scale cultivation and analysis

*100g* system #1 design and installation: The *100g* system was designed and constructed as described by Flory et al.[18] out of 80/20 T-slot framing to support a motor-driven chain circulating a sorbent belt and attached to a 4.2 $m^2$ raceway pond. In the first design, ~256 sorbent packets, each containing ~ 2 g of A501 sorbent beads, were attached to parallel chains made of Acetal (ServoCity) plastic (to prevent rusting and metal leaching into the pond) to form a belt of sorbent packets. The chains of the belt were attached to coaxially mounted sprockets connected by a shaft, including one driven by a single servo motor to control the belt speed. The belt was slowly rotated to expose the packets to the air for capturing $CO_2$ and to the pond for moisture-driven $CO_2$ release into the alkaline pond medium.

*100g* system #2 design and installation: A second *100g* system was built with two independent loops (see Figure 3 of Flory et al.[18]), each with independent drive motors to provide redundancy in case one of them might fail (e.g., if the chain disconnected). The size of the frame was expanded to accommodate ~375 two-gram sorbent packets (nearly twice that of system #1).

Pond-scale abiotic $CO_2$ capture and delivery evaluation: The *100g* system was installed in a 4.2 m$^2$ pond (Commercial Algae Professionals), equipped with a YSI 5200A-DC (YSI Inc., Yellow Springs, OH, USA) water quality monitoring system that simultaneously measured pH, pond water temperature (°C), dissolved oxygen saturation (%), and salinity (g/L) as described by McGowen et al..[22] The pond was retrofitted to replace the paddlewheel with a centrifugal pump to minimize passive $CO_2$ capture independent of the DAC system and evaluated for abiotic $CO_2$ delivery as described by Flory et al..[18]

Cultivation in 4.2 m$^2$ raceway ponds: Wild-type S*ynechocystis* 6803 as well as *Synechocystis* TE/Δ*slr1609* were scaled at ASU's Arizona Center for Algae Technology and Innovation (AzCATI) following their standard methods from flasks to indoor photobioreactors to 110 L flat panels in a greenhouse using BG-11 medium modified to contain a 16:1 N:P molar ratio with 5 mM $NaNO_3$ and 35 mM $NaHCO_3$. A similar protocol was used to scale *Chlorella vulgaris* 1201 with a starting pH of ~8.5 that rose to ~10 during cultivation.

### 3.4 Used-sorbent packet washing and analysis

Used-packet washing: Polymer packets recovered from the *100g* system and unused (new) packets were evaluated for abiotic $CO_2$-release capacity. Upon receipt, packets were dried overnight in a laminar-flow hood. Then packets were submerged in 1 L of 0.1% (v/v) peracetic acid while being stirred at 100 rpm for 1 h, after which packets were rinsed under a stream of DI water for 30 min and agitated in 2 L of DI water for another 30 min. Packet were then transferred to 1 L of 0.04% (w/v) sodium hypochlorite, stirred at 100 rpm for 30 min, rinsed again under running DI water for 10 min, and stirred once more in 2 L of DI water for 30 min. The packets were next subjected to an acid wash by stirring overnight (~14 h) in ~1 mM HCl (pH 3.0), followed by a 10-min running DI water rinse and a 30-min stir in 2 L DI water. For bicarbonate exchange, packets were soaked in 0.5 M $NaHCO_3$ (1 L) for 12 h, while being stirred at 100 rpm, transferred to fresh 0.5 M $NaHCO_3$ (1 L) for an additional 24 h, rinsed once more under running DI water for 10 min, and finally stirred in 1 L DI water for 30 min. After air-drying in the laminar hood (> 12 h), the washed packets were immediately subjected to the Abiotic test described above to evaluate the sorbents $CO_2$ capacity and time dependent release; $CO_2$ was reported as the amount (mg) of $CO_2$ released per packet (2 g of polymer).

Periodic acid-Schiff (PAS) stain: For the PAS stain, new A501 sorbent beads and A501 sorbent beads that had been used in the *100g* system and those that had been used in the *100g* system but were subsequently washed in 1 mM HCl (mesh were cut into small pieces) were oxidized by incubation in 0.5% (w/v) periodic acid solution for five minutes in the dark. Following three ~1-minute rinses with DI water, samples were stained by incubation in Schiff's reagent for 15 minutes in the dark. After staining, the samples were rinsed in DI water for five minutes. In the PAS reaction, periodic acid oxidizes vicinal diols in polysaccharides like glycogen and in mucosubstances like glycoproteins and glycolipids. This oxidation cleaves the bond between two adjacent carbons that are not involved in the glycosidic linkage or ring closure within the monosaccharide units of long polysaccharides. As a result, aldehyde groups are formed at the termini of each broken monosaccharide ring, which then react with Schiff's reagent, producing a purple-magenta color. A variation of the used packet washing protocol described above was used without peracetic acid to wash packets before repeating the PAS stain.

Estimating the number of used sorbent packets, wet/dry cycles, cultivation days, and cumulative UV exposure: We hypothesize wet/dry-cycle-induced swelling/shrinking mechanical stresses, exposure to microalgae and cyanobacteria, and cumulative UV exposure are important contributors to sorbent and mesh degradation observed during outdoor trials. To quantify the exposure to these stressors, first the number of wet/dry cycles was determined for sorbent packets used during outdoor trials for 35, 119, 190, 217, 262 and 302 days from March 28, 2024 to February 17, 2025. For initial trials, the length of a wet/dry cycle was fixed at two hours during the day with higher average wind speed and temperature, and at four hours at night due to lower average wind and temperature. To optimize $CO_2$ delivery, the belt rotation software was updated on April 29, 2024 to advance the packets in segments only when the next segment of packets was assumed to be 90% loaded with $CO_2$ based on the actual measured wind speed and a model predicting $CO_2$ loading from data recorded in a wind tunnel at various wind speeds.[18] The cumulative time each packet was exposed to the air and water was logged each day. The main point of uncertainty was determining how many cycles occurred when the system was down, since only the date and not the exact time of failure and restarting was recorded. Thus, for the days when the system was down, a low estimate of zero wet-dry cycles and a high estimate based on the system operating the full day were assigned.

Next, the number of days the sorbent packets were exposed to microalgae or cyanobacteria during cultivation trials was determined based on the operating log from the day the culture was inoculated in the pond (typically done early in the morning but sometimes not until noon) until the cultivation trial was stopped and the pond was drained (typically done in the morning but sometimes not until early afternoon). Finally, the cumulative UV exposure for each packet was determined from the historical monthly average maximum UV index during each operating month as sourced from OpenWeatherMap, converting the UV index (UVI) to MJ m$^{-2}$ (1 UVI sustained for 1 hour = 0.025 MJ m$^{-2}$) and then normalizing the UV intensity for each hour of the day by multiplying the UVI by $\sin(\beta)$ where $\beta$ is $2\pi / \tau$ and $\tau$ is the number of daylight hours predicted for that particular day as sourced from OpenWeatherMap. Since 10% of circulation path of the packets was underneath the rest of the packets that could partially shade the packets, we determined a low estimate that 10% of the time the packets received no UV exposure (when partially covered ) and a high estimate assuming full UV exposure during the entire wet/dry cycle.

Thermogravimetric analysis (TGA): The TGA measurements were carried out on a TA Instruments TGA Q50 ramping from 40 to 100 ºC and holding isothermally for 10 min to ensure the removal of moisture, cooling down to 40 ºC, and a second ramp from 40 ºC to 500 °C at a heating rate of 10 °C min$^{-1}$ under $N_2$ atmosphere. The samples were completely dried prior to the analysis.

Attenuated Total Reflectance Fourier Transform Infrared Spectroscopy (FTIR-ATR): ATR-FTIR spectra (400−4000 cm$^{-1}$) were obtained with a Nicolet iS50 FTIR spectrometer at 4 cm$^{-1}$ resolution and averaged over 64 scans. Samples were dried under a fume hood overnight prior to testing.

Sorbent particle image analysis: Particle shape and size metrics were determined using optical microscopy with a Morphologi G3SE Particle Analyzer. Sorbent was carefully dispersed onto a large glass plate on the sample stage to minimize aggregates. Using a 2.5x lens, a rectangular area of 132 x 84 mm was scanned, and images of all particles were captured over ~ 1 hour. Two focal points were used in the image capture to ensure the full distribution of particle sizes was obtained. The Morphologi software calculated shape and size metrics, which were exported and run through an in-house MATLAB script to visualize final figures and tables. The primary size metric used to describe each particle distribution is the volume-weighted average diameter ($d_{4,3}$), which reduces skew compared to number-based averages of distributions and better represents the bulk when the number of particles measured is limited.

Sorbent particle true density analysis: Sorbent volumes were measured using an AccuPyc II 1340 Helium Gas Pycnometer (Micromeritics, Norcross, GA, USA) with a set pressure of 19.5 psig. Skeletal volume was determined from 10 cycles of pressurization-depressurization with helium gas while using a 1 cm$^3$ sample cup. Skeletal density was calculated as the volume divided by the mass of sorbent in the sample cup during the 10 cycles and averaged. The material's skeletal volume excluded open pores of a particle and void space between particles but included closed pores or other spaces that may be inaccessible by helium at 19.5 psig.

Sorbent tensile strength analysis: A series of diametric compression tests (Brazilian indirect tension tests) were conducted on unused (new) A501 beads, as well as beads that had been used for outdoor trials for 35, 190, and 302 days. These tests were performed using a TA Instruments Hybrid Discovery Rheometer 2 operating in displacement-control mode with a loading rate of 1 μm/s. The maximum displacement was set to 50% of the bead's diameter to ensure complete failure. The diameter of the bead was measured using the rheometer by noting the distance between the compression plates when a force larger than 0 mN was first detected as the plates were brought together with the bead between them. All tests measured force and displacement over time and were carried out under ambient conditions (20 °C and ~25-30% RH). For each condition, 2–5 replicates were tested, with fewer replicates on the longer-used beads. It was difficult to find undamaged samples suitable for testing in the 302-day-used condition. Moreover, there were challenges in handling and mounting the highly fragile, heavily used beads. From the measured maximum force, the tensile strength of each sample was calculated using the equation proposed by Hiramatsu and Oka:[23]

$$S_t = 0.9 \frac{P}{4R^2} \qquad (Eq. 1)$$

where P is the maximum axial force and R is the radius of the bead. This equation was based on observations from compression testing of several kinds of irregularly shaped rocks as well as theoretical calculations of stress within a linear elastic spherical specimen subject to diametric compression and has been widely used to approximate the strength of beads of various materials.[24–26]

### 3.5 Commercial Feasibility Analysis

Technoeconomic analysis and life cycle assessment (TEA/LCA). This analysis is built on a modular engineering process model that integrates the mass and energy flows across a co-located algal cultivation and biorefining system. The model simulates the entire process chain from biomass production in open raceway ponds, integrating the described moisture-driven $CO_2$ DAC and delivery system through downstream conversion that includes the recovery of high-value phycocyanin and the remaining biomass converted to fuels via hydrothermal liquefaction (HTL). The model describing the DAC process is provided in Flory et al.,[18] the cultivation subprocess model is described in Davis et al.[27] and the hydrothermal liquefaction subprocess model is described in Chen et al.[28] This foundational modeling provides the infrastructure for concurrent TEA and LCA.

Economic performance was assessed using a discounted cash flow rate of return (DCFROR) method to determine the minimum fuel selling price (MFSP) that achieves a net present value of zero over a 30-year facility lifetime. The analysis assumes "nth-of-a-kind" economics, including a 10% internal rate of return and a 35% tax rate. Additional financial assumptions, such as depreciation schedules, debt structure, and plant start-up periods, follow established conventions.

Capital and operating costs were derived by scaling unit operations based on process flow rates and system throughput using exponential scaling factors. Equipment sizing for cultivation and dewatering systems was based on maximum volumetric and mass flow requirements. Conversion infrastructure, including HTL and fuel upgrading systems, was sized using annual average flow and energy demand, with seasonal variability addressed through intermediate biomass storage. Cost data were sourced from literature and normalized to 2021 dollars using the Chemical Engineering Plant Cost Index. Variable costs, including nutrients, hydrogen, and natural gas, were similarly adjusted. Process model outputs were used to calculate annual production volumes and determine the MFSP.

An attributional life cycle assessment was performed to quantify the environmental impacts of the modeled system. The system boundary spans from microalgal cultivation, dewatering, biomass storage, conversion, product distribution, to end use. Results are presented for global warming potential (GWP) using a 100-year time horizon. All impacts were normalized to a functional unit of 1 MJ of liquid fuel, accounting for the combined energy content of renewable diesel and naphtha products. Life cycle inventory (LCI) data were derived from Ecoinvent 3.7 and analyzed using OpenLCA 1.11. Greenhouse gas emissions were characterized using IPCC factors of 1, 29.8, and 273 for $CO_2$, $CH_4$, and $N_2O$, respectively. Emission factors for indirect inputs were sourced from the GREET model and supporting literature.

## 4. Results and Discussion

Sorbent biocompatibility and flask-scale cultivation. Four commercially available strong-base anion exchange resins (AER)—three beaded AERs including HPR-4800 (DuPont), Amberlite IRA-900 (DuPont) and A501 (Purolite) and one sheet AER, Excellion (Snopure)—were evaluated in terms of their ion exchange capacity (IEC), $CO_2$ delivery kinetics into aqueous medium and biocompatibility. The measured IEC, which is theoretically twice its maximum $CO_2$ capture capacity, was found to be between 2.1 and 2.4 mmol per gram for all beaded AERs (**Table S1**). The IEC of Excellion sheets was lower due to structural materials used to form the sheets and reported to be 1.2 mmol per gram.[29] The $CO_2$ delivery capacity and kinetics of each sorbent were determined by submerging the dry sorbent loaded with $CO_2$ from ambient air into a solution of 10 mM $Na_2CO_3$ and converting the change in solution pH to the amount of $CO_2$ released following the methods described by Shesh et al.[21] **Figure S1A** shows the $CO_2$ release kinetics and capacities of each of the evaluated sorbents; they were similar for the three beaded sorbents whereas the Excellion sheets were 3–4 fold slower as previously reported by Flory et al.[18] Interestingly, the HPR-4800 sorbent showed markedly reduced capacity during the third run that persisted thereafter (**Figure S1B**). We hypothesize that HPR-4800 has some weakly bound quaternary ammonium functional groups that are lost under repeated or prolonged (several hours) exposure to alkaline solutions (pH 10–11). Xia et al. showed that AEMs are susceptible to several degradation pathways when exposed to alkaline conditions, including Hofmann elimination, nucleophilic substitution, and ylide formation.[30]

In order to evaluate the biocompatibility, 1 g of each beaded sorbent was immersed in a 50 mL culture of a mutant strain of *Synechocystis* 6803 modified to secrete laurate (*Synechocystis* TE/Δ*slr1609*)[19] and supplemented with 20 mM $NaHCO_3$ added each day as a carbon source. As shown in **Figure S2**, the cultures with new A501 and new IRA-900 had a brief lag phase before growing rapidly to an optical density at 730 nm ($OD_{730}$) of 4.2 and 3, respectively, after 5 days, which was significantly higher than a control culture that was given 20 mM bicarbonate daily, indicating that the sorbents provided additional $CO_2$ to enhance the growth. In addition, the pH of the cultures containing sorbent remained lower than the pH of the control culture, which typically reached 11 by day three when supplied with 20 mM bicarbonate; at this pH growth slows down. On the other hand, the density of cultures exposed to new HPR-4800 sorbent declined significantly after 48 hours, indicating that this polymer appears to be less biocompatible than the others. **Figure S3** shows photographs of the cultures, with healthy cultures showing a dark green color due to substantive pigment production and unhealthy cultures showing a blueish or reddish color, or clear, indicating a lack of biomass.

Quaternary ammonium (QA or $NR_4^+$)-containing polymers are known to have antimicrobial properties that are dependent on the chain length connecting the QA sites.[31,32] We hypothesized that some of the reduced growth was due to weakly bound polymer fragments that were not stable in the alkaline medium (pH 9.5–10), causing fragments of AERs containing multiple quaternary ammonium cations ($NR_4^+$) to be released into the growth medium, where they may intercalate and disrupt *Synechocystis* cell membranes. Thus, the biocompatibility test was repeated using the same sorbent: if weakly bound fragments are released upon first use, used sorbent may be more biocompatible. As shown in **Figure S2**, the growth of all cultures exposed to used sorbents improved significantly for the first ~three days, and cultures exposed to used HPR-4800 did much better than those exposed to the new polymer. However, the cultures grown with used IRA-900 began to decline after three days. Cultures grown with A501 did well with both new and used polymer, suggesting that this was the most bicompatible of the three.

A subsequent six-day cultivation study with used IRA-900 showed that the sorbent particles fractured into small fragments that were able to escape containment through the 25-μm pores of the mesh bag (**Figure S4**), so this sorbent was not evaluated further. HPR-4800 was also not evaluated further due to the decline in performance after initial use and apparent instability under alkaline conditions. Both Excellion and A501 AERs appeared to be stable under alkaline conditions; this may be the main reason they are biocompatible. Other studies have also shown that AERs, including those with QA groups, can be biocompatible, including when directly contacting cells.[33,34] Although Excellion was found to be biocompatible (**Figure S5**) we did not continue evaluating it in this study due to significantly slower $CO_2$ delivery kinetics compared to A501 (**Figure S1A**).

Next, the most promising AER, A501—which showed good biocompatibility and alkaline stability after multiple incubations with *Synechocystis* TE/Δ*slr1609* cultures, good IEC and $CO_2$-binding capacity, and rapid $CO_2$ delivery kinetics into an alkaline medium—was used to cultivate *Synechocystis* TE/Δ*slr1609* without $NaHCO_3$ supplementation to evaluate the sorbent's ability to deliver $CO_2$ to support growth. The sorbent, previously used for cultivation with *Synechocystis* TE/Δ*slr1609* either once or twice, was immersed each day for 30 minutes in the cultivation medium to release $CO_2$ and then left to dry and load with $CO_2$ from ambient air flow in a laminar flow hood overnight; the process was repeated each day for four days. As shown in **Figure 2**, the culture grew well, faster than the 20 mM $NaHCO_3$ control, to an $OD_{730}$ of about 3.2 after four days. This corresponds to a growth rate of about 190 mg dry weight $L^{-1}$ $d^{-1}$, and the additional accumulated biomass corresponds to 68–76 mg of $CO_2$ delivered by A501 to the 50 mL of culture over 4 days. The observed growth rates are consistent with the control before it slowed due to high pH, but is still lower than for wild-type *Synechocystis* 6803 grown with bubbled $CO_2$ and with optimal media and light conditions.[35,36]

In order to assess the impact of the cyanobacterium on the $CO_2$ capacity of the A501 sorbent, at the conclusion of the cultivation trial, the $CO_2$ delivery capacity of A501 was assessed by submerging the dry sorbent into a solution of 10 mM $Na_2CO_3$ as previously described. As shown in in **Table S2**, used A501 had about a third the $CO_2$ delivery capacity of new A501 (Figure S1), and the capacity did not change much between the first and second use. Since A501 contains $NR_4^+$, we hypothesized that some of the $HCO_3^-$, $CO_3^{2-}$ and $OH^-$ anions required for moisture-driven $CO_2$ capture were exchanged with other anions in the BG-11 growth medium, including nitrate ($NO_3^-$), chloride ($Cl^-$), sulphate ($SO_4^{2-}$) and phosphate ($HPO_4^{2-}$). For example, $NO_3^-$ is present at high concentration (17.6 mM) in standard BG-11 medium, much more than is needed, and may bind to the resin, release $HCO_3^-$ into the medium, and reduce the sorbent's capacity to capture additional $CO_2$ during subsequent cycles. Thus, modified BG-11 medium was prepared in which the $NO_3^-$ levels were reduced by half to 8.8 mM, and 35 mM $NaHCO_3$ was added to ensure that ~80% of the anions in solution (and in equilibrium with the AER) are available to support moisture-driven DAC. **Figure S6** shows the $CO_2$ released by a mesh packet containing 2 g of A501 into 0.5 L of 20 mM $Na_2CO_3$ before and after being immersed in modified BG-11 medium with 50% nitrate and 35 mM $NaHCO_3$. Although the $CO_2$ delivery kinetics were reduced ~three-fold, the sorbent was still able to deliver within one hour ~90% of the amount of $CO_2$ delivered by sorbent that was freshly exchanged with $NaHCO_3$.

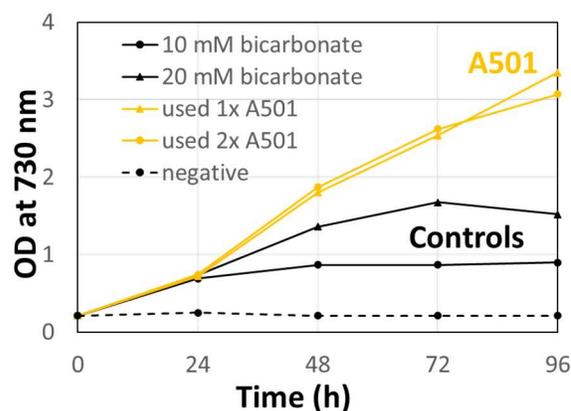

**Figure 2.** Growth of *Synechocystis* TE/Δ*slr1609* cultures in BG-11 medium with daily immersion for 30 minutes of A501 sorbent, used once or twice previously for cultivation, followed by drying in ambient air, compared to positive controls with 10 or 20 mM $NaHCO_3$ added once and a negative control without any $NaHCO_3$.

<u>Bench-scale cultivation trials using the *1g* system in a laminar flow hood</u>. In order to evaluate the A501 sorbents' ability to continuously deliver $CO_2$ to cultivate cyanobacteria, 1 g of A501 sorbent beads were loaded into each of 16 tubular mesh packets and attached to a conveyance system described by Flory et al.[18] for circulating the sorbent between ambient air and the alkaline growth medium and installed within a laminar flow hood. The initial design capacity was to deliver ~1 g $CO_2$ per day so we call this the *1g* system.[18] The *1g* system was started with 12 L of fresh modified BG-11 medium (50% nitrate) with 17.5 mM $Na_2CO_3$ without cells to allow the *1g* system to fill the buffer with $CO_2$ captured from air and produce $HCO_3^-$ in preparation for cultivation. The pH and inorganic carbon levels were monitored over time. The initial trial showed excellent $CO_2$ delivery (6 g) over the first 3 d (**Figure S7B**). However, after the medium was inoculated with *Synechocystis* TE/Δ*slr1609*, the cells grew slowly until we increased the light intensity from 50 to 250 µmol photons $m^{-2} s^{-1}$ on day 8, which significantly increased growth (**Figure S7A**); the pH of the medium also increased to pH 10 by day 12, which is typical for well-growing cultures (**Figure S7B**). The light intensity has been shown to have a significant impact on the growth rate of *Synechocystis* 6803.[36]

Another cultivation trial was started in fresh modified BG-11 medium with a light intensity of 250 µmol photons $m^{-2} s^{-1}$ from the start of the experiment. The pH changes and $CO_2$ released into the *1g* system over 72 h with the A501 belt are shown in **Figure S8**. The pH in the medium dropped from 10.5 to about 9.1 within 72 h, which corresponds to about 6 g of $CO_2$ released by the belt into the medium in three days, similar to the first trial. *Synechocystis* TE/Δ*slr1609* cultivation was started after 72 h at a starting optical density (OD) of 0.2 at a light intensity of 250 µmol photons $m^{-2} s^{-1}$. As shown in **Figure 3A**, growth of the *Synechocystis* TE/Δ*slr1609* culture in the *1g* system took off immediately and the optical density doubled in 24 h. The density of the *Synechocystis* TE/Δ*slr1609* culture in the *1g* system continued to increase and reached $OD_{730}$ of 1.3 by day 8, and then growth slowed. Addition of an extra dose of nitrate did not help. The pH changes in the medium over 10 days are shown in **Figure 3B**. After cell cultivation was started on day 3, the pH in the medium increased as the culture grew and reached pH 10.2 by day 7. After day 7 the pH in the medium started dropping suggesting $CO_2$ delivered by the *1g* system outpaced $CO_2$ fixation by *Synechocystis* TE/Δ*slr1609*.

The net $CO_2$ delivered by the *1g* system was determined by the sum of the inorganic carbon (IC) delivered and retained in the 12 L of medium by the *1g* system (**Table S3**) and the $CO_2$ consumed (fixed) by *Synechocystis*

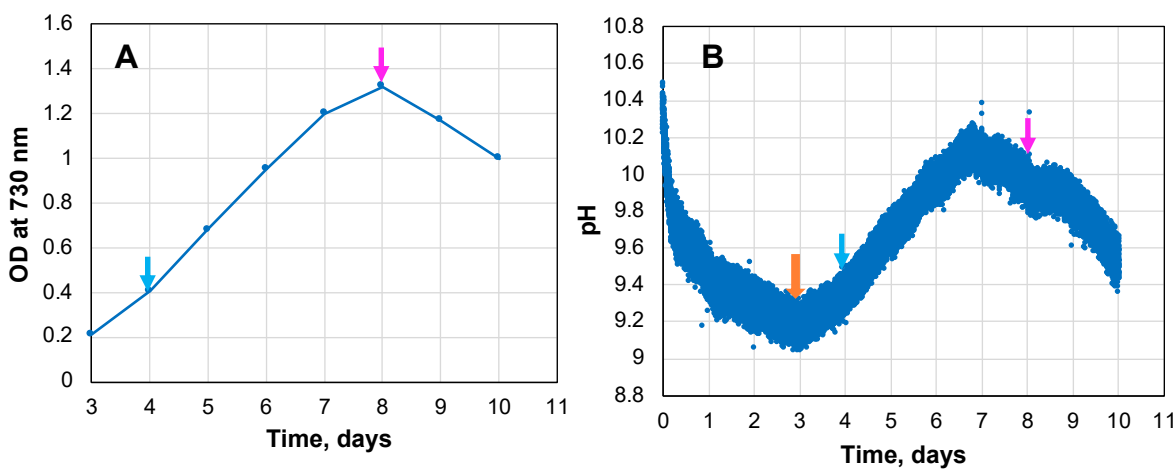

**Figure 3**. Growth (A) and pH changes (B) in 12 L of *Synechocystis* culture in the *1g* system at a light intensity of 250 μmol photons m$^{-2}$ s$^{-1}$. Orange arrow indicates the start of cultivation of *Synechocystis*; blue arrows indicate addition of an extra dose of phosphate; pink arrows indicate addition of an extra dose of nitrate.

TE/Δ*slr1609* from the accumulated biomass in the medium shown in **Figure 4** and tabulated in **Table S4**. **Figure 4** shows a clear transition from $CO_2$ delivered by the *1g* system initially to $CO_2$ being fixed by *Synechocystis* TE/Δ*slr1609* cells leading to increased biomass. Prior to cultivation, the amount of $CO_2$ released by the A501 belt into 12 L of medium in the first three days increased over time and reached about 6 g in total by day 3, which is consistent with that calculated based on the pH change. After cultivation was started, the amount of $CO_2$ in the medium started dropping when *Synechocystis* TE/Δ*slr1609* started growing and consuming (fixing) more $CO_2$ than was being delivered by the *1g* system. The IC data showed that $CO_2$ released by 16 g of sorbent in the *1g* system was sufficient for vigorous growth of *Synechocystis* TE/Δ*slr1609* (39 mg L$^{-1}$ d$^{-1}$). In the first three days the belt delivered on average 2 g of $CO_2$ per day, and the total amount of carbon in the culture increased further by ~0.5 g $CO_2$ per day on average in the presence of cells, which is consistent with the observed increase in biomass of ~0.5 g, indicating the growth rate was carbon limited. This suggested the presence of the cells does reduce performance of the sorbent, but the *Synechocystis* TE/Δ*slr1609* culture looked healthy throughout the cultivation trial as shown in **Figure S9**. This is consistent with the flask scale cultivation trials showing reduced time-dependent $CO_2$ release from the sorbent after use.

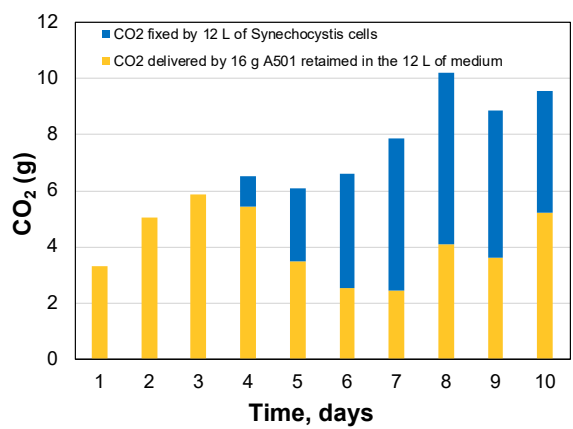

**Figure 4.** Total $CO_2$ delivered to 12 L *Synechocystis* TE/Δ*slr1609* culture by 16 g A501 over 10 days. Air/belt-delivered $CO_2$ retained in 12 L of medium (**yellow bars**) and $CO_2$ fixed by *Synechocystis* cells (**blue bars**).

Outdoor cultivation trial in 4.2 m² raceway ponds. In order to evaluate the moisture-driven DAC technology under more realistic operating conditions, the conveyance system design was scaled up 100-fold to a *100g* system designed to deliver ~100 g $CO_2$ per day as described by Flory et al..[18] Approximately 250 packets each containing 2 g of sorbent were attached to a single chain. The system was attached to a 4.2 m² pond at ASU's Arizona Center for Algae Technology and Innovation (AzCATI) that had the paddlewheel replaced by a centrifugal pump to minimize air exchange and $CO_2$ delivery directly from ambient air and not via the sorbent in the *100g* system. The initial trial planned for fall 2023 was delayed due to technical issues so the pond was inoculated with wild type *Synechocystis* 6803 in December 2023 (**Figure 5**) and compared to a control pond without the *100g* system.

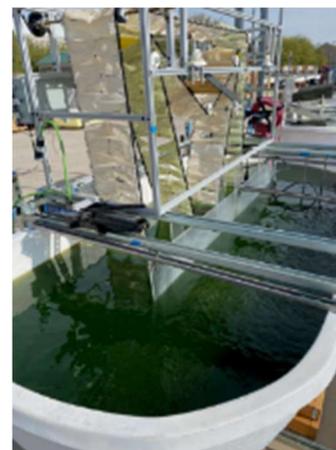

**Figure 5.** First *100g* system with a single long loop used for a cultivation trial in December 2023.

Although the cold temperatures (5–20 °C) limited the growth of this mesophilic strain in both ponds to 2.5–2.8 g biomass per m² pond surface area per day (**Figure S10**), the *100g* pond showed no contamination, while the control pond became contaminated with amoeba and diatoms (**Figure S11**). The clean culture from the *100g* pond was used to inoculate the control pond a second time after cleaning, and again, no contamination was observed in the *100g* pond, while the control pond became contaminated (**Figure S12**). We hypothesize that a small amount of quaternary ammonium ($NR_4^+$)-containing polymer fragments may be released by the sorbent that are tolerated by the cyanobacterium but that inhibit otherwise slow growth of contaminants during a cold period that generally has low contamination pressure.

The *100g* system was then redesigned with two independent loops (see Figure 3 of Flory et al.[18]) to **1)** allow at least one loop to run should the other malfunction and **2)** expand capacity to include 375 two-gram sorbent packets (nearly twice that of system #1). This system ran much more reliably than the first-generation system. In April 2024, *100g* system #2 was installed at AzCATI, and it was found to deliver ~100 g $CO_2$ per day.[18] Unfortunately, we had difficulty scaling the *Synechocystis* TE/Δ*slr1609* strain, so the pond was inoculated at a low (0.1) $OD_{730}$, resulting in the culture dying within a few days (**Figure S13**). *100g* was run for an additional week to see if the culture might recover. The *100g* pond began accumulating IC (**Figure S14**) indicating the *100g* system was still actively delivering $CO_2$. Interestingly, on June 13 a green alga *Desmodesmus* sp. blew into the pond and began growing. The green alga's growth plateaued around $OD_{730}$ 0.3 so the culture was split into a second pond that was supplemented with $CO_2$ with a pH setpoint of 10 to match the *100g* system to see if the slow grown may be due to carbon limitation. The culture with $CO_2$ supplementation grew to higher OD than the one in the *100g* pond and the IC levels in the *100g* declined throughout the period when *Demodesmus* was growing suggesting growth in the *100g* pond was carbon limited. We hypothesized the sorbent packet $CO_2$ delivery performance was reduced from visual biofouling by the green alga.

Two additional trials with the *100g* system were performed from September 25 to November 25 (fall) 2024 including 27 days of cultivation with the green alga *Chlorella vulgaris* 1201 and December 19, 2025 to February 17, 2025 (winter) 2025 including 33 days of cultivation using *Synechocystis* 6803. During the fall 2024 trial, 118 new sorbent packets were installed with 182 used packets on the *100g* system, which delivered ~100 g $CO_2$ per day (**Figure S15**) or about 0.17 g $CO_2$ per g sorbent per day, in line with the expected performance of the combination of new and used packets installed. However, as soon as the pond was inoculated with *Chlorella vulgaris* 1201, the $CO_2$ delivery performance of the packets became negligible (**Figure S15**). During the winter 2025 trial, although the *100g* system pond maintained a somewhat higher level of inorganic carbon (**Figure S16**) and lower pH (**Figure S17**) than the control pond without the *100g* system before and during cultivation with *Synechocystis* 6803, the amount of $CO_2$ delivered (2–10 g $CO_2$ per day) was substantially lower than anticipated and did not lead to significant differences in growth compared to the control pond (**Figure S18**). This indicated a significant performance degradation of the sorbent when immersed in cyanobacteria that warranted further investigation.

Sorbent packet biofouling, cleaning and performance loss. We hypothesized that exopolysaccharides secreted by microalgae and cyanobacteria into the cultivation medium were fouling and reducing the $CO_2$ delivery performance of the sorbent packets. Sorbent packets used for the first two cultivation trials with *Synechocystis* TE/Δ*slr1609* and *Desmodesmus* were analyzed using a periodic acid-Schiff (PAS) stain protocol that oxidizes like glycogen and mucosubstances like glycoproteins and glycolipids, which react with the Schiff's reagent producing a purple-magenta color. As shown in **Figure 6**, the fouled A501 sorbent beads became stained with a deep magenta color whereas the new beads were not stained. The used beads were then washed in 1 mM HCl (pH ~3) overnight to breakdown and remove the polysaccharides. As shown in **Figure 6C**, the washed beads had significantly less stain. **Figure S19** shows an electron micrograph of a new and used sorbent bead, where the used bead appears to have less porosity and a smoother surface consistent with fouling. The mesh packets also appeared to have clogged pores under the light microscope (**Figure S20A, B**) compared to a new packet (**Figure S20C**) and after washing with 1 mM HCl (pH ~3) also showed significantly less fouling (**Figure S20D**). Li et al. also reported fouling of AERs by protein-like substances from exposure to municipal wastewater.[37]

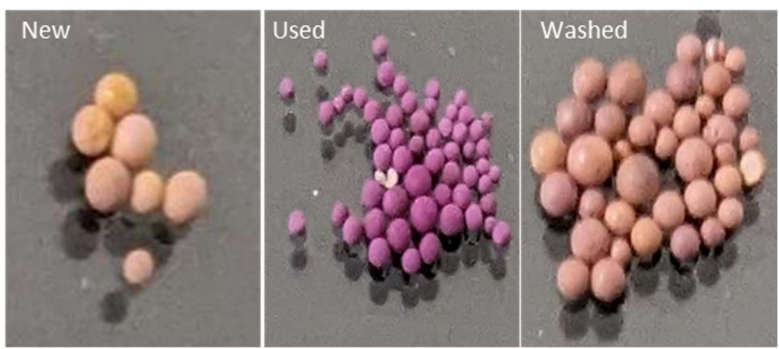

**Figure 6:** Photographs of new, used (in *100g* system) and washed (with 1 mM HCl, pH 3.0) A501 polymers after staining with periodic acid-Schiff stain.

To address biomass fouling on the mesh from exopolysaccharides, a wash protocol was developed. Due to their extensive use in water filtration, anion exchange membranes (AEM) biofouling and cleaning protocols have been well documented in the literature and were reviewed by Merino-Garcia and Velizarov.[38] The most common treatments are washing with acid or base, enzymes, surfactants, $H_2O_2$, $O_3$, EDTA, membrane and ion exchange with NaCl. Sodium hypochlorite (NaClO) has also been used extensively, though it can degrade the AEM after prolonged exposure, including cracking, reduced conductivity, loss of quaternary ammonium functional groups and even the polystyrene-divinylbenzene backbone.[39,40] These studies saw the worst effects at higher concentrations (5000 ppm). The A501 manufacturer (Purolite) also recommends washing A501 in 0.1% (v/v) peracetic acid.

Sorbent packets used during outdoor trials for 0 days (new), 35 days and 302 days (including 0, 32 and 78 days of cultivation with cyanobacteria and microalgae; see **Table S5**) were washed in 0.1% (v/v) peracetic acid for 1 h, followed by 0.04% (w/v) or 400 ppm sodium hypochlorite for 0.5 h, 1 mM HCl (pH ~3) solution overnight, then ion exchanged twice overnight in a 0.5 M $NaHCO_3$ solution to remove the $Cl^-$ ions introduced during the washing process, rinsed in DI water and finally air dried overnight to load with $CO_2$. The time-dependent $CO_2$ release into a 10 mM $Na_2CO_3$ solution of the packets was measured before and after the cleaning protocol. As shown in **Figure 7**, the $CO_2$ binding capacity of the packet used for 35 days was reduced to ~25% of its initial capacity; the cleaning protocol was able to restore to ~70% of its initial capacity but with markedly slower kinetics, suggesting that the cleaning protocol may not have been able to remove biofouling from the interior pores of the microporous sorbent beads. On the other hand, the $CO_2$ binding capacity of the packet used for 302 days was reduced to ~10% of its initial capacity and the cleaning protocol was only able to restore ~35% of its initial capacity suggesting that some of the capacity loss may be permanent due to loss of $NR_4^+$ $CO_2$ binding sites. Sorbent used for 0, 35, 190 and 302 days were also transferred to new mesh packets to evaluate the impact of mesh degradation or clogged pores the cleaning protocol and/or $CO_2$ delivery. As shown in **Figure S21A**, the performance of the cleaned sorbent declined with use with a significant decline between 190 and 302 days. The new mesh also led to faster $CO_2$ delivery, as illustrated in **Figure S21B** for the cleaned sorbent used for 35 days in its original mesh vs a new mesh packet.

<u>Mesh Packet Damage Analysis</u>. To investigate causes of reduced performance of the *100g* system during the cultivation trials, all 280 of the mesh packets remaining on the *100g* system at the conclusion of ~9 months of outdoor use, including four cultivation trials, were removed and sorted qualitatively according to the visual wear, damage and sorbent loss. Most of the 180 packets that were installed on October 21, 2024 and January 13, 2025 and operated during the milder fall/winter months (October 2024 to February 2025) had no significant damage to the mesh packet, with only three having minor sorbent loss. Whereas the 100 packets installed on March 28 and July 18, 2024 and operated during the Spring/summer (Mar to July 2024) with additional exposure to the hot temperatures and high UV intensity during the harsh summer months saw some damage to ~90% of the packets (**Figure S22**). About 96% of the packet damage was attributed to mesh degradation-related damage (tears to side or end seams) with the remaining (4%) due to mechanical stress-related damage to the mesh (torn at tabs or attachment points; **Figure S22**). Of these 100 older sorbent packets, only ~10% had no sorbent loss, ~60% had minor sorbent loss (quantified as more than ~70% sorbent remaining), while ~30% of the packets had significant sorbent loss (quantified as less than ~70% sorbent remaining). We hypothesize that significantly higher UV irradiation and temperatures during the summer months weakened the heat-sealed seams containing the sorbent beads within the mesh packets leading to damage and sorbent loss. This damage could potentially be mitigated using UV-stabilized mesh materials and ultrasonic welding to seal the mesh together.

Loss of quaternary ammonium $CO_2$ binding sites with use. Changes to the chemical composition of the beads that may be responsible for the loss of $CO_2$ capacity in used sorbents were analyzed by Attenuated Total Reflectance Fourier Transform Infrared Spectroscopy (ATR-FTIR) and thermogravimetric analysis (TGA). From the chemical composition reported by the manufacturere (Purolite),[41] A501 has a crosslinked polystyrene backbone functionalized with quaternary ammonium (QA) groups. It has been reported that the stability of QA containing polymers under alkaline conditions can undergo different degradation pathways depending on the valence electrons. Velazco-Medel et al.,[42] reported styrene-based polyionic liquids with QA and a $HCO_3^-$ counterion were found to undergo oxidative degradation after four weeks of wet-dry cycles due to air exposure during sorbent drying and $CO_2$ loading from ambient air at a faster rate compared to with a halide ($Cl^-$) counterion. Other potential degradation pathways include nucleophilic attack to the α-carbon next to the positively charged nitrogen and Hofmann elimination of β-carbons by acidic hydrogens. **Figure S23A** shows the ATR-FTIR spectra for a new A501 sorbent as it comes from the manufacturer with a $Cl^-$ counterion (inactive for $CO_2$ capture), after ion exchange with $NaHCO_3$, and used for outdoor trials for 0 (new), 190, and 302 days. The data show considerable changes to the chemical composition of the resin; in particular, the QA bands at 850, 1380, and 1620 cm$^{-1}$ decrease their intensity and shift after 190 and 302 days of use, indicating a decrease in QA concentration. The loss of QA is further confirmed by TGA (**Figure S23B**), where the thermal degradation of the QA around 170 ºC is only significant for unused sorbent, but is significantly reduced in the sorbent used for 190 and 302 days. This loss of QA groups may be driven by repeated wet-dry cycling, regular exposure to alkaline solutions (pH ~10), photosynthetic microorganisms, UV light and oxygen in the air.

Changes to sorbent size, shape and tensile strength. In addition to sorbent loss attributed performance reduction and chemical changes, we also investigated the physical integrity of the sorbent after use in the *100g* system for 0, 35, 190 and 302 days using optical microscopy with particle size distribution analysis, scanning electron microscopy (SEM) and rheometry. Throughout the course of the 302 days of outdoor use, new mesh packets containing new sorbent beads were installed periodically to replace mesh packets that became physically damaged during the trial. The conditions subjected to the packets installed on four different dates throughout the trial are summarized in **Table S5**, including the install date, operating dates, number of days used, number of days exposed to and species of photosynthetic microorganism during cultivation trials, estimated number of cumulative wet/dry cycles, and estimated cumulative UV exposure. **Figure 8** shows optical images of the sorbent beads after 0, 35, 119, 190, 217, 262 and 302 days of use and **Figure 9** shows image analysis of the particles after 0, 35, 190 and 302 days of use.

New sorbent beads have very high circularity and aspect ratio (**Figure 9**) with an average volume-weighted diameter ($d_{4,3}$) of 723 μm ± 440 μm (1 standard deviation), which is at the middle of the manufacturer's specified range of 425–1200 μm. The vast majority of the sorbent particles used only for the final 35 days of the trials, during the cool months of January and February 2025 with very low UV exposure and 35 days exposed to the cyanobacterium *Synechocystis* TE/Δ*slr1609*, appeared unchanged with only a few beads showing some evidence of cracking and fracturing, which reduced the aspect ratio and circularity slightly; the particle diameter decreased somewhat to 699 μm ± 411 μm that is likely due to the small sample size in the image analysis and is not a statistically significant difference. The sorbent particles used for 119 days, with an additional 85 days used during the warm fall months and moderate UV exposure, and an additional 31 days for cultivation with the green alga *Chlorella vulgaris* 1201, showed more smaller sorbent fragments, but generally were still mostly spherical in appearance (**Figure 8**). The sorbent particles used for 190 days, with an additional 71 days of abiotic use during the hot summer months but no additional exposure to photosynthetic microorganisms, showed an increase in particle fragmentation (**Figure 9b**), with a slight increase in volume-weighted average diameter to 718 ± 455 μm.

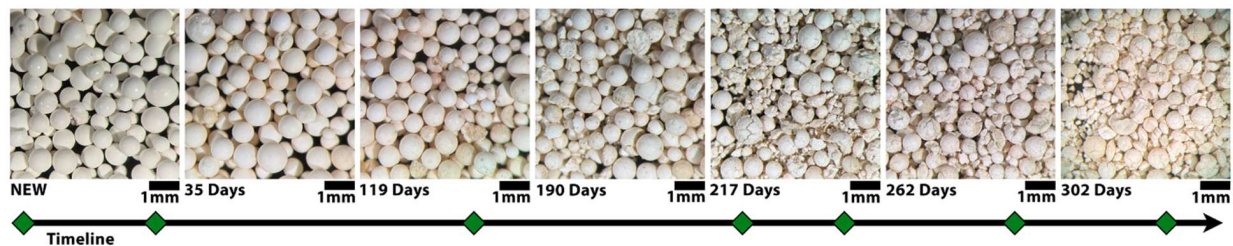

**Figure 8.** Optical micrographs of A501 sorbent beads exposed to outdoor wet/dry cycles for 0 to 302 days.

The largest reduction in sorbent aspect ratio and circularity occurred in the time from 35 to 190 days of use. Since aspect ratio is a measure of the length to width of the sorbent particles, spherical particles have values closer to one. At 190 days, the aspect ratio decreased more than the circularity compared to 35 days, indicating that circular particles have degraded into more oblong and elongated shapes, further supporting the particle fragmentation observed in the size distribution. The sorbent particles used for 217 days, including an additional 27 days of use of which eight were exposed to the green microalgae *Desmodesmus* showed significantly higher number of sorbent fragments and cracked beads (**Figure 8**). The beads exposed to 262 and 302 days, with an additional 45 and 85 days of use, respectively, and not exposed to any additional cultivation trials, were also significantly fragmented and cracked with an average volume-weighted diameter of 646 μm ± 420 μm for those used for 302 days, but not much different than the beads used for 217 days that had ~30% fewer total wet/dry cycles but the same number of days used for cultivating photosynthetic microorganisms.

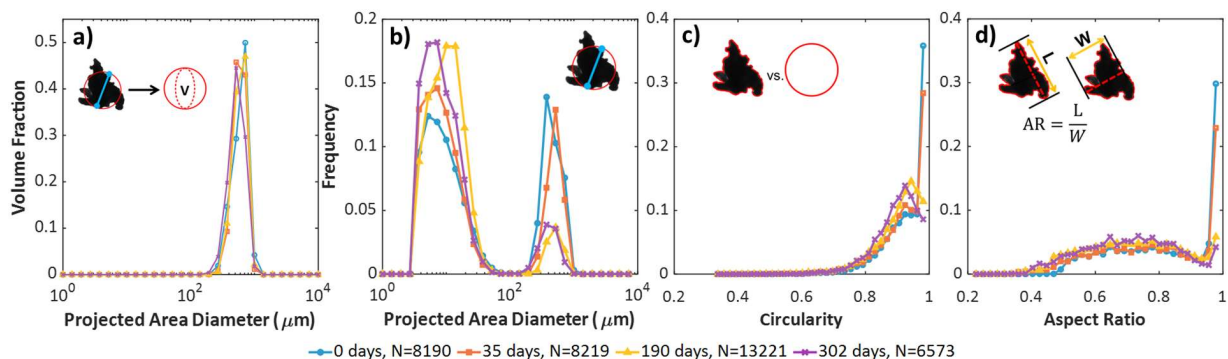

**Figure 9.** Changes in sorbent bead diameter on a) a volume basis and b) a number basis, as well as shape through c) circularity and d) aspect ratio at four time points of days used.

The reduction in particle size is also supported by the true density increasing from 1.28 ± 0.02 to 7.10 ± 0.02 g/cm$^3$ from day 0 to 302, where smaller particle fragments pack in between the larger particles to significantly increase the mass per volume. Scanning electron microscopy (SEM) images (**Figure S24**) of single beads used for 0, 190, and 302 days did not show any significant changes to the microstructure and porosity of the sorbent, but they decreased their dimensions considerably and sorbent beads used for 302 days showed some evidence of fouling on the surface. The lack of change in porosity is further supported by the small difference in true volume from 0.19 ± 0.01 to 0.17 ± 0.03 cm$^3$ from day 0 to 302, respectively.

The tensile strength of individual sorbent beads was also analysed with a rheometer by compressing individual beads until failure. **Figure 10** illustrates how the tensile strength of the A501 beads declines substantially with use. In this figure, the error bars show one standard deviation from the mean strength for each use condition, based on multiple replicate tests. These bars provide a measure of the variability in the data and indicate the consistency of the tensile strength values within each group. Notably, the strength of the sorbent used for 302 days is almost one-quarter that of the new sorbent. We hypothesize that the primary driver of mechanical fracturing was due to internal stresses generated from a significant moisture gradient across the bead during each wet/dry cycle that was exacerbated from biofouling-induced reduced water transport. As the tensile strength of the sorbent decreases with use it eventually cannot withstand the internal stress from the water gradient and fractures.

These results suggest that avoiding direct contact of the sorbent with the microalgae or cyanobacteria, such as by immersing the liquid in the media recycle stream after biomass harvest, may be critical to prolonging the life and $CO_2$ delivery performance of the sorbent bead during repeated wet-dry cycling expected during commercial outdoor use. Alternatively, Ghoussoub reported poly(4-styrenesulfonate)-based antifouling coatings that significantly reduce biofouling in a green alga *Chlamydomonas reinhardtii* compared to the uncoated resin while consuming less than 0.1% of their ion exchange capacity.[43]

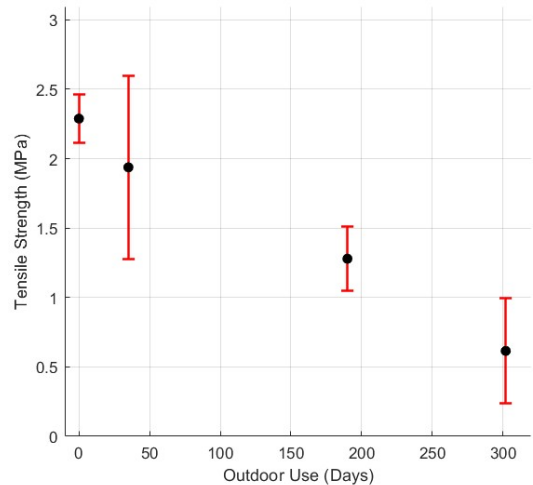

**Figure 10.** Changes in sorbent tensile strength with use in outdoor trials.

*Commercial feasibility and environmental impacts*. The costs and greenhouse gas emissions associated with an $n^{th}$-of-a-kind biorefinery composed of moisture-driven DAC for use in cultivating microalgae and cyanobacteria were assessed. This assessment assumes the performance of the sorbent measured in the laboratory and outdoors prior to direct contact with microalgae or cyanobacteria, which can be achieved by interfacing the sorbent with the media recycle stream to reduce biofouling-related sorbent performance degradation, and UV-stabilized mesh materials can extend their lifetime from ~6 months to 5 years. The engineering process model is shown in **Figure 11**.

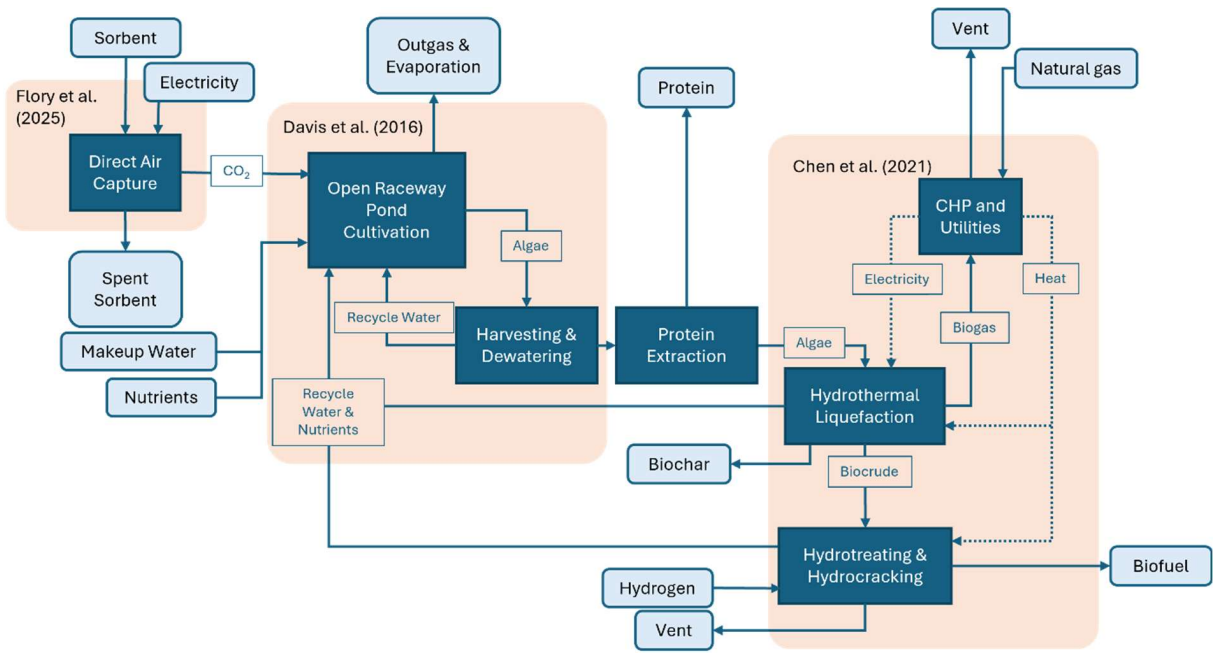

**Figure 11.** Engineering process model of moisture-driven DAC and delivery to cultivate cyanobacteria with phycocyanin and other protein extraction to provide high and mid-value co-products, followed by hydrothermal liquefaction of the residual biomass to biofuel and biochar.

The DAC subprocess model is described in Flory et al.,[18] the cultivation subprocess model is described in Davis et al.[27] and the hydrothermal liquefaction subprocess model is described in Chen et al.[28] Additional modifications include extracting the protein fraction of the cyanobacterial biomass to be sold as protein supplements and animal feed, and the phycocyanin protein fraction extracted and sold as a natural blue pigment.

Microalgae cultivated with $CO_2$ supplied by conventional DAC are currently cost-prohibitive. Therefore, the cost of cultivating microalgae is typically assessed assuming $CO_2$ will be supplied from a point source at a very low cost (~$50/tonne). This, however, severely limits the production capacity of microalgae cultivated in this manner given the mismatch between $CO_2$ sources and good growing regions for microalgae. Previous studies have predicted microalgae to cost $510/tonne ash-free dry weight, but this is assuming a point source supplied $CO_2$ is delivered with a high utilization efficiency (**Figure 12**).[27,44] In reality, $CO_2$ utilization efficiencies are still lower (40%) than the 90% target, increasing the cost to $650/tonne. To achieve the large capacity of algae cultivation required to satisfy aviation fuel demand or to achieve food security, DAC $CO_2$ sources are needed, but current costs for DAC would increase costs further to >$1,100/tonne. However, preliminary technoeconomic analyses (TEA) show that moisture-driven DAC and delivery for cultivating cyanobacteria can be used to reduce the cost to $500/tonne through increased utilization efficiency and the potential for $CO_2$ costs lower than $51/tonne.

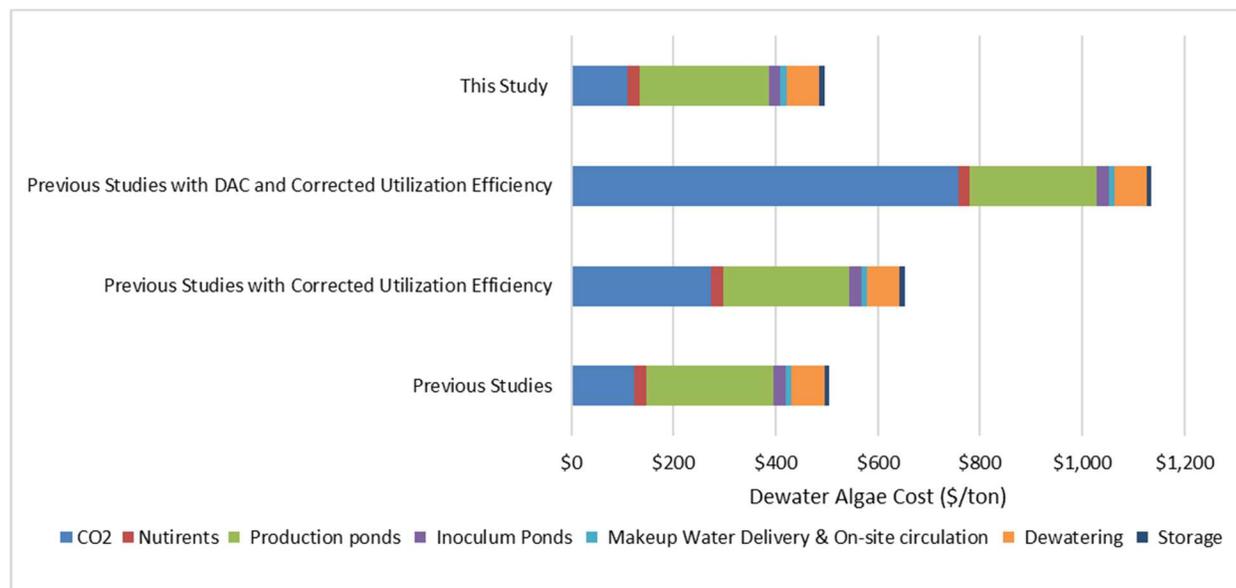

**Figure 12.** Comparison of cost to produce dewatered algal biomass based on 90% carbon utilization efficiency (CUE; previous studies), a correction assuming more realistic CUE of 40%, additional costs for sourcing $CO_2$ from DAC using the same more realistic CUE, and the current study using moisture-driven DAC for $CO_2$ delivery.

Microalgae are a promising feedstock for renewable fuel, but even with improvements in $CO_2$ delivery, renewable fuel from microalgae is currently not cost-competitive with fossil fuels. Renewable fuels from microalgae tend to cost >$7.00/gasoline gallon equivalent (GGE; **Figure 13**). However, preliminary TEA shows that high-value revenue generating cyanobacterial coproducts can be used to sell cost-competitive fuels ($2.50/GGE). Using the model from Chen et al.[28] HTL can produce fuel at $2.50/GGE when biomass feedstock is provided free of charge. Selling phycocyanin and other high-value products offsets the cultivation and biomass processing costs, providing feedstock free of charge or at a steep discount for HTL. Preliminary TEA shows that to provide biomass free of charge to HTL, the co-product revenue for protein, assuming 40% protein in the cyanobacteria (Spirulina is 60–70%), needs to be at least $1.4/kg of protein. Assuming a market value for protein of $6/kg (the value of whey protein used as a human heath supplement) that leaves $4.2/kg for the extraction process which was not modelled here but would consist of cell disruption and separation. The value of protein and Phycocyanin vary by market from $150/kg for analytic grade tracer to $1/kg animal feed supplements. Unfortunately, high value markets are unable to support large scale fuel production, but with a market size in the hundreds of thousands of tonnes in the United States for protein as a human health supplement, fuel production at a competitive price point could exceed 500 barrels per day of gasoline equivalents, which is large enough to constitute a small first-of-a-kind biorefinery.

Preliminary life cycle analysis (LCA) shows the carbon intensity of the biofuel produced is 33 g$CO_{2e}$ MJ$^{-1}$ using a cutoff methodology that assigns the burden of algae cultivation to the protein product, which is significantly less than 88 g$CO_{2e}$ MJ$^{-1}$ for traditional fossil-based diesel. Another important aspect of the LCA that was not examined here is the potential contamination of the biomass and protein products with compounds released from the degrading sorbent that could reduce their value and make them unsuitable for some markets (e.g., food and beverage). However, for the process to be economical, the sorbent degradation must be very slow, such that it maintains its performance for 3–5 years, and thus the amount of sorbent compounds released into the products must inherently be very low, such that they may be below the detection limit.

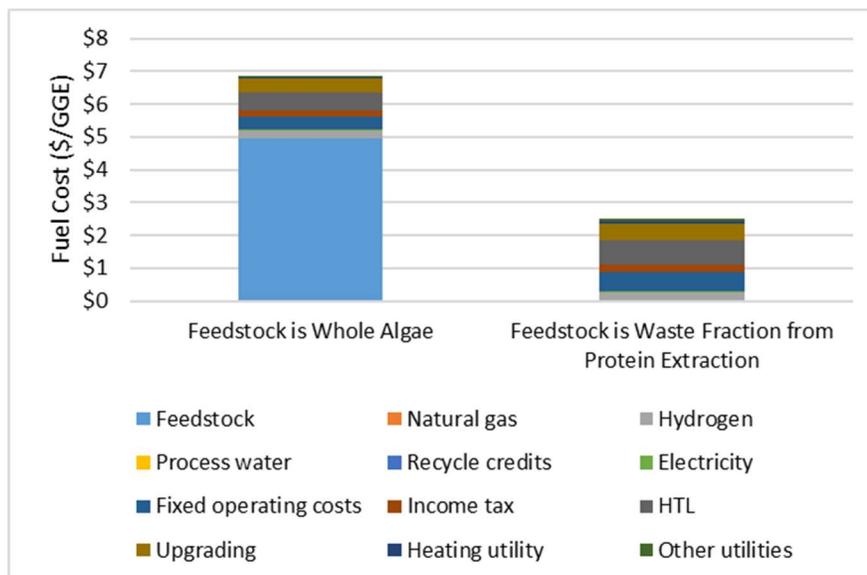

**Figure 13.** Cost analysis of processing whole biomass to biofuels compared to extracting phycocyanin (natural blue colorant) and other proteins before converting the residual biomass to biofuel.

## 5. Conclusion

Moisture-driven $CO_2$ capture and delivery into alkaline solutions has significant potential to reduce energy requirements for DAC over processes and sorbents that use heat to release captured $CO_2$, as well as $CO_2$ losses from sparging concentrated $CO_2$ into shallow raceway ponds. This could enable low-cost $CO_2$ capture and delivery at the site where it is utilized for cultivating cyanobacteria and microalgae. Further research is needed to develop sorbents that resist biofouling and sorbent degradation from prolonged direct and repeated contact with cyanobacteria and microalgae with hundreds of days of outdoor exposure and wet-dry cycling to prevent significant reductions in the sorbents $CO_2$ binding capacity and delivery kinetics. Indirect methods of interfacing the sorbents with growth media recycle streams may be required to retain DAC sorbent performance. Further, research is also needed to develop new AER materials that are tolerant of prolonged exposure to alkaline conditions and that reduce the magnitude and duration of moisture gradients induced mechanical stresses from cyclic water loading as well as to understand the composition, quantity and toxicity of any compounds released by the sorbent as it ages and the potential impact on product quality and market requirements. Extracting high-value phycocyanin (natural blue food and beverage dye) and other proteins from cyanobacterial biomass prior to conversion to biofuel opens up a pathway to produce a small biorefinery (500 barrels per day) of cost-competitive biofuel ($2.50/GGE). Taken together, with further development moisture-driven DAC processes can play an important role in increasing domestic production of carbon feedstock sourced at the site of cultivation for photosynthetic microbes that produce bio-based fuels, chemicals and materials and support a net zero carbon economy.

## 6. Acknowledgements

This material is based upon work primarily supported by the U.S. Department of Energy's Office of Energy Efficiency and Renewable Energy (EERE) under the Bioenergy Technology Office award number DE-EE0009274 including design, data collection and analysis. Data analysis and manuscript preparation was also supported by the U.S. Department of Energy, Office of Science, Office of Basic Energy Sciences under Award Number DE-SC0023343. We would also like to thank Ani Nazari for her help in measuring the ion exchange capacities of the beaded sorbents.

## 7. Author Contributions

- **Justin Flory**: conceptualization, data curation, formal analysis, funding acquisition, methodology, project administration, supervision, visualization, writing – original draft, writing – review & editing
- **Shuqin Li**: investigation ($CO_2$ delivery, biocompatibility and cell culturing), data curation, formal analysis, methodology, visualization, writing – review & editing
- **Samantha Taylor**: investigation (system design, build, operation), data curation, formal analysis, methodology, supervision, visualization, writing – review & editing
- **Sunil Tiwari**: investigation ($CO_2$ delivery), data curation, formal analysis, methodology, visualization, writing – review & editing
- **Garrett Cole**: investigation (technoeconomic and life cycle analyses), data curation, formal analysis, methodology, visualization, writing – review & editing
- **Marlene Velazco Medel**: investigation (ATR-FTIR, TGA), data curation, formal analysis, methodology, visualization, writing – review & editing
- **Amory Lowe**: data curation, investigation (control system, $CO_2$ loading model), software, visualization, writing – review & editing
- **Jordan Monroe**: investigation (particle analysis), writing – review & editing
- **Sara Sarbaz**: investigation (tensile strength analysis), writing – review & editing
- **Nick Lowery**: investigation (system design, build, operation), writing – review & editing
- **Joel Eliston**: investigation (system design, build, operation), writing – review & editing
- **Heidi P. Feigenbaum**: supervision (tensile strength analysis), writing – review & editing
- **Heather Emady**: supervision (particle analysis), writing – review & editing
- **Jason C. Quinn**: conceptualization, formal analysis (TEA/LCA), funding acquisition, methodology, supervision, writing – review & editing
- **John McGowen**: conceptualization, funding acquisition, methodology, supervision (outdoor trials), writing – review & editing
- **Matthew Green**: conceptualization, formal analysis, funding acquisition, methodology, supervision (sorbent analysis), writing – review & editing
- **Klaus Lackner**: conceptualization, formal analysis, funding acquisition, methodology, supervision (sorbent analysis), writing – review & editing
- **Wim Vermaas**: conceptualization, formal analysis, funding acquisition, methodology, project administration, supervision (indoor cultivation), writing – review & editing

## 8. Conflicts of Interest

The authors declare no conflicts of interest related to the content of this paper. M.D.G. is co-founder of NuAria, LLC and owns 50% equity interest in the company. The work of this manuscript is not directly related to the activities of the company. K.S.L is a paid advisor to DACLab, a DAC company, and Aircela Inc., which aims to produce fuel from air and renewable energy and owns small stakes in those companies. Arizona State University has licensed part of its DAC intellectual property to Carbon Collect Limited and owns a stake in the company. As an employee of the University, K.S.L. is a technical advisor to the company and in recognition also received shares from the company. Carbon Collect Limited also supports DAC research at Arizona State University. K.S.L. also has a financial stake in Aircela and is on the Board of Aircela. None of these companies are involved in the work reported here.

## 9. Supplementary Information

**Table S1** summarizes the data used to determine the ion exchange capacities of several commercial AERs measured using Mohr's method including the mass of the polymer (AER sorbent), volume of 0.1 M $AgNO_3$ titrant added when the color change occurred, the number of moles of chloride ($Cl^-$) released into solution based on the quantity of titrant added, which is assumed to be equal to the number of accessible quaternary ammonium (QA) sites in the AER that can be exchanged with $NaHCO_3$, and the corresponding ion exchange capacity (IEC). The measured IECs are similar between the different sorbents and within the expected ranges provided by the manufacturer's data sheets.

**Table S1**: Summary of the titration results to determine the IEC of several beaded AERs.

| Sample Name | Mass of Polymer [g] | Volume of 0.1M $AgNO_3$ [ml] | Moles of $Cl^-$ | Moles of QA | Measured IEC [mmol/g] |
|---|---|---|---|---|---|
| Amberlite HPR4800 | 2.01 | 48.3 | 0.00483 | 0.00483 | 2.3 |
| Purolite A-501P | 2.05 | 41.7 | 0.00417 | 0.00417 | 2.1 |
| Amberlite IRA900 | 2.00 | 48.10 | 0.00481 | 0.00481 | 2.4 |

**Figure S1A** shows the $CO_2$ release from four different AER sorbents into a 10 mM $Na_2CO_3$ solution. The beaded sorbents (A501, IRA-900, HPR-4800) are all 1 g with similar IECs (see Table S1) and the Excellion sheet was 8" x 0.75" in size (1.76 g) with a manufacturer specified IEC of 2.8–3.6 milliequivalents. All beaded sorbents released most of their $CO_2$ within 15–20 minutes whereas the Excellion sheet took closer to 1 hour. All samples were run in triplicate except HRP-4800, which is shown in duplicate since the performance after the third run declined substantially and the performance loss persisted in subsequent two runs as shown in **Figure S1B**.

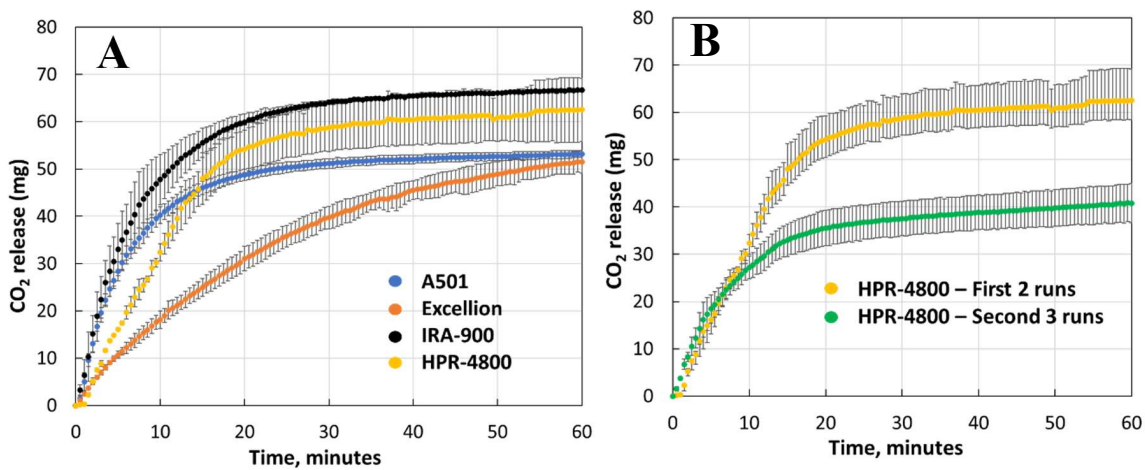

**Figure S1. (a)** $CO_2$ released by A501, Excellion, IRA-900 and HPR-4800 AERs into 10 mM $Na_2CO_3$. **(b)** $CO_2$ released by HPR-4800 comparing the first two and second three runs.

**Figure S2** shows the growth of *Synechocystis* TE/Δ*slr1609* in the presence of HPR-4800 (DuPont) Amberlite IRA-900 (DuPont) and A501 (Purolite) strong base AERs continually immersed in the medium and addition of 20 mM NaHCO$_3$ each day as the carbon source compared to a positive control with 20 mM NaHCO$_3$ added daily and no sorbent and a negative control with no NaHCO$_3$ supplementation. HPR-4800 new was run as two independent replicates. The IRA-900 cultures turned blueish after 48 hours indicating the cultures were producing less chlorophyll and not healthy. **Figure S3** shows images of the cultures taken on day 3, 5 and 7.

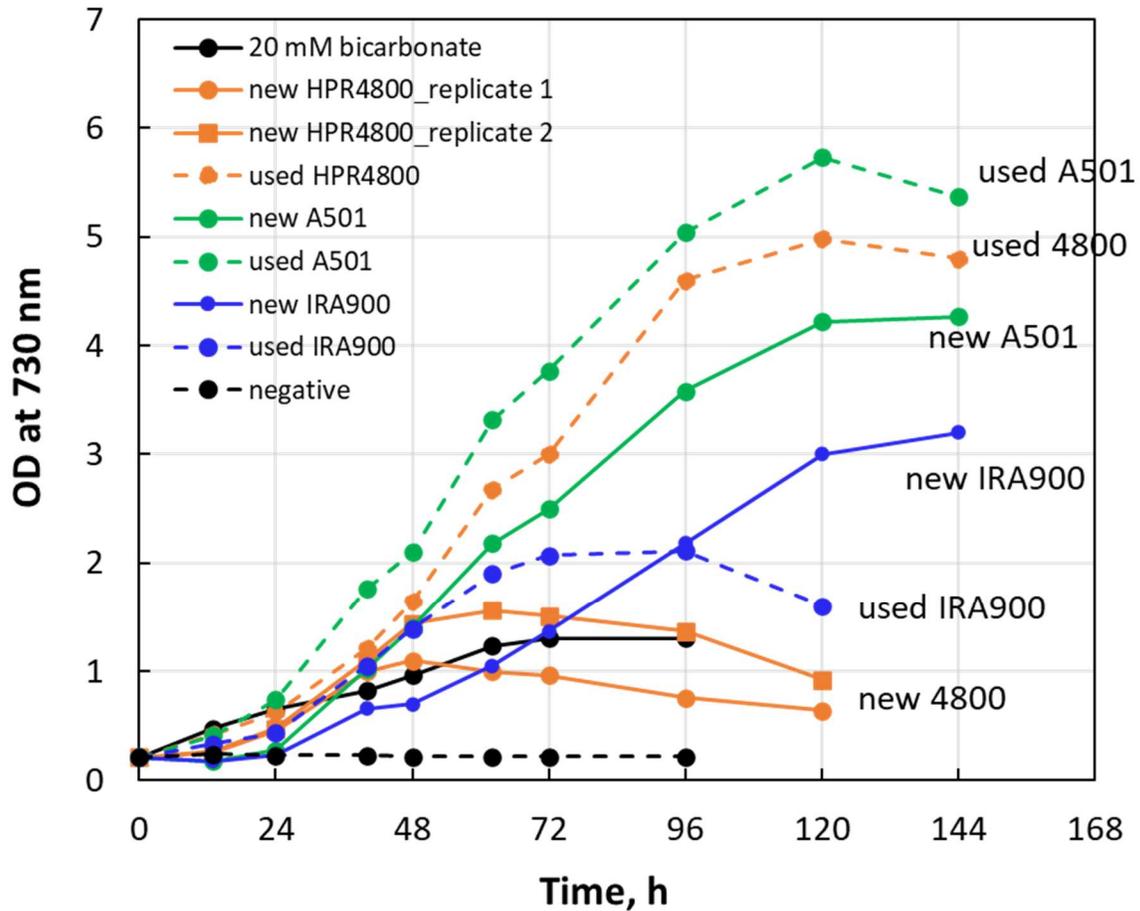

**Figure S2.** Growth of TE/Δ*slr1609* in the presence of HPR-4800 (DuPont) Amberlite IRA-900 (DuPont) and A501 (Purolite) comparing the first use of the sorbent (new) and the second use (used).

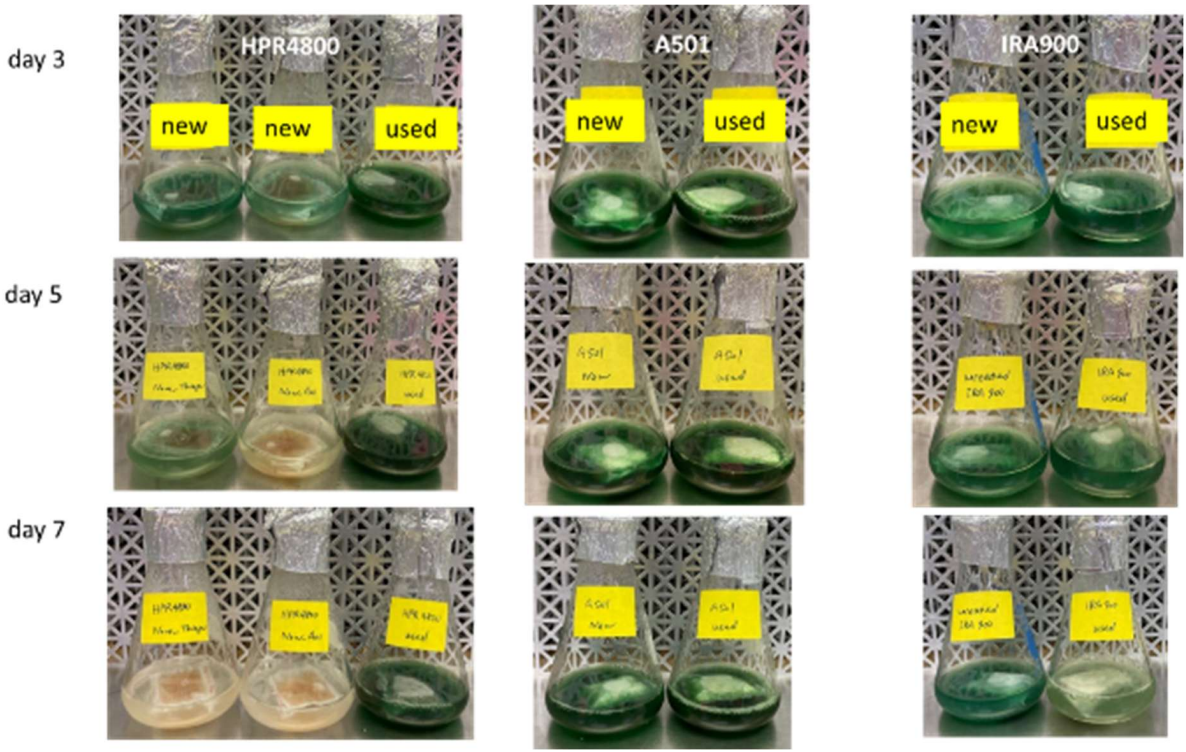

**Figure S3.** Photographs of the *Synechocystis* TE/Δ*slr1609* cultures that were grown and exposed to new or used HPR-4800, A501 and IRA-900 commercial AERs for 3, 5 and 7 days (top, middle and bottom, respectively).

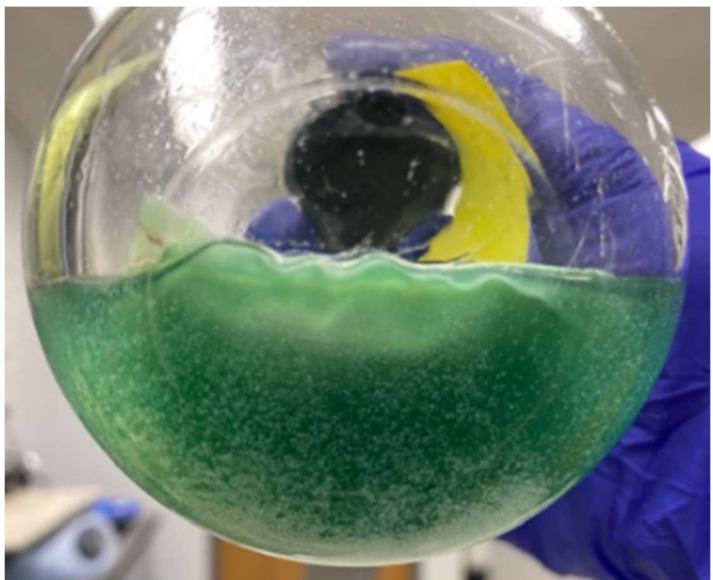

**Figure S4.** Photograph showing a six-day old *Synechocystis* TE/Δ*slr1609* culture incubated with IRA-900 enclosed in a mesh bag with 25 μm pores. Note that the IRA-900 fragments have escaped from the mesh bag into the medium.

**Figure S5** shows growth of *Synechocystis* TE/Δ*slr1609* cultures in 50 ml BG-11 medium with daily immersion for 30 minutes of Excellion sorbent, indicated by red arrows, followed by drying in ambient air until the next immersion the following day, compared to positive controls with 10 or 20 mM NaHCO$_3$ added once and a negative control without added NaHCO$_3$.

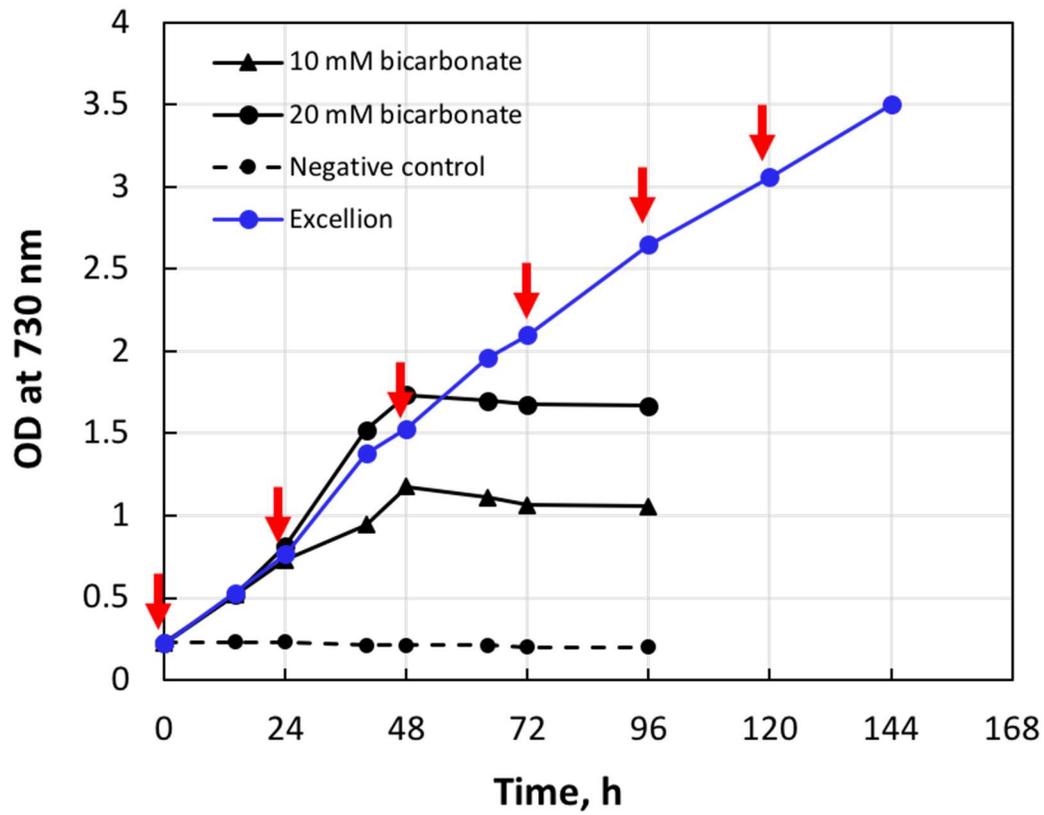

**Table S2.** Inorganic carbon released by A501 into a 30 ml 10 mM Na$_2$CO$_3$ solution after being immersed in standard BG-11 medium used for cultivation with *Synechocystis* TE/Δ*slr1609* cultures for 30 minutes each usage.

| Sample | CO$_2$ released (mg) | Bicarbonate in 30 ml (mM) |
|---|---|---|
| 1x used A501 | 15.4 | 11.7 |
| 2x used A501 | 16.9 | 12.8 |

**Figure S6** shows the CO$_2$ delivery kinetics of a mesh packet containing A501 after being immersed in the modified BG-11 medium for eight hours.

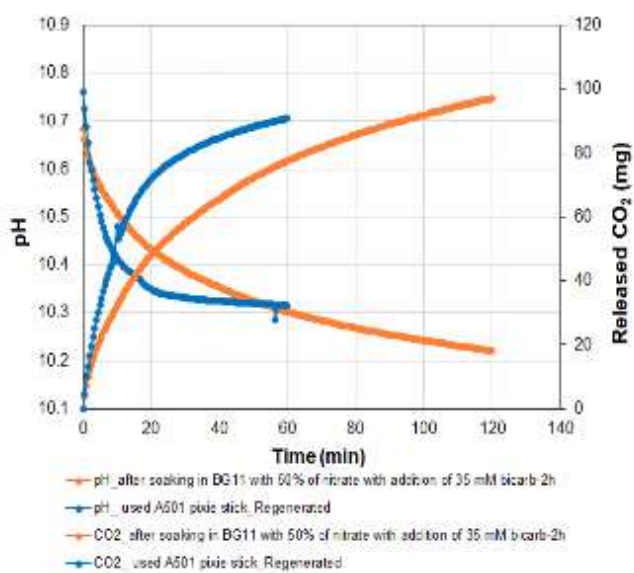

**Figure S6.** CO$_2$ release kinetics from an A501 mesh packets wih 2 g of sorbent after regeneration (blue) followed by an eight-hour soak in BG11 (50% nitrate) with 35 mM bicarbonate (orange).

**Figure S7** shows cultivation of *Synechocystis* TE/Δ*slr1609* grown with a light intensity of 50 μmol photons m$^{-2}$ s$^{-1}$. In spite of substantial carbon in the medium after growth slowed after 1 day. Addition of extra phosphate helped a little, but increasing the light intensity from 50 to 250 μmol photons m$^{-2}$ s$^{-1}$ on day 8 significantly increased growth (**Figure S7A**) and the pH of the medium increased to pH 10 by day 12, which is typical for well-growing cultures (**Figure S7B**).

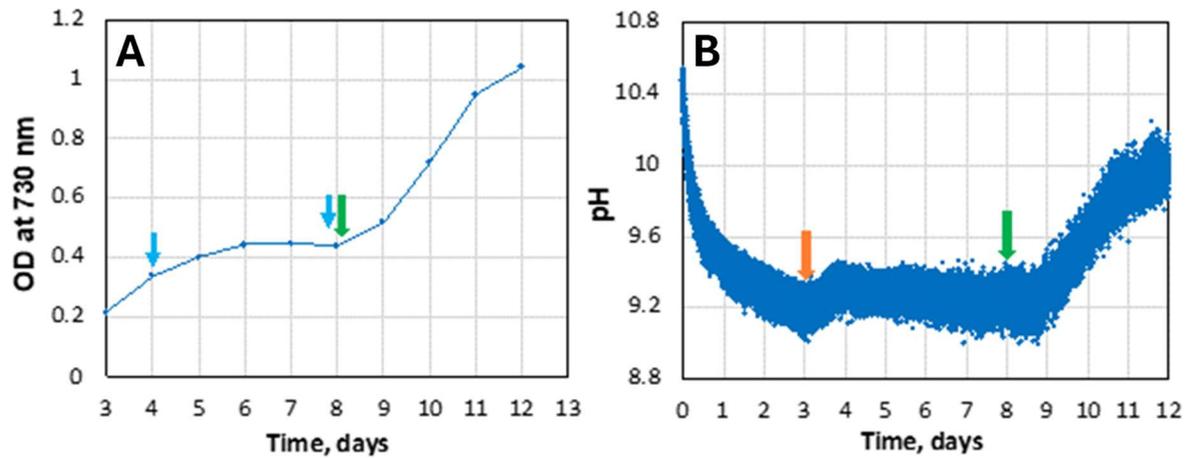

**Figure S7.** Growth of *Synechocystis* (**A**) and pH changes (**B**) in 12-liter BG11 (50% nitrate) with 17.5 mM Na$_2$CO$_3$ supported by the A501 belt on a 4-h cycle; *Synechocystis* cultivation started on day 3 (orange arrow). The light intensity from day 3 until day 7 was 50 μmol photons m$^{-2}$ s$^{-1}$ and was increased to 250 μmol photons m$^{-2}$ s$^{-1}$ on day 8 (green arrow). Extra phosphate was provided on days 4 and 8 (blue arrows).

**Figure S8** shows the pH change due to corresponding mass of CO$_2$ released by the A501 belt into the medium prior to inoculation with *Synechocystis* TE/Δ*slr1609*

**Table S3. Inorganic carbon (IC) of the BG-11 medium during the 2$^{nd}$ *1g* trial with *Synechocystis* TE/Δ*slr1609*.** IC data on day 0 is from the added 17.5 mM Na$_2$CO$_3$ and sample was taken before running the A501 belt. The net CO$_2$ released by the A501 belt into the 12 L of medium was calculated by subtracting the IC from day 0. Total net CO$_2$ delivered by 16 g A501, including IC retained in the 12 L of medium as well as the CO$_2$ fixed by *Synechocystis* TE/Δ*slr1609* cells over 10 days, is listed in **Table S4**. CO$_2$ fixed by 12 L of *Synechocystis* TE/Δ*slr1609* cells in the *1g* system was calculated based on the biomass.

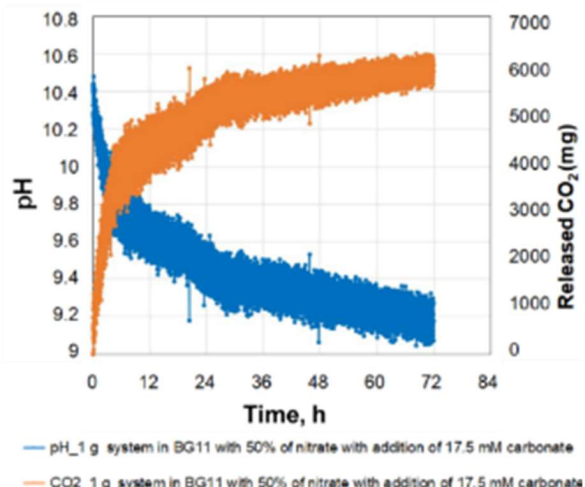

**Figure S8**. pH changes and deduced CO$_2$ release from the regenerated A501 belt to 12 L of BG11 medium (50% of nitrate) with 17.5 mM Na$_2$CO$_3$ prior to adding cells.

**Table S3.** Raw IC data in the 12 L of medium in the *1g* system over 10 days.

| Sample | Raw IC data (mM) | Net IC delivered by A501 belt (mM) | Net CO$_2$ delivered by A501 belt (g) |
|---|---|---|---|
| day 0 | 20.0 | | |
| day 1 | 26.3 | 6.3 | 3.3 |
| day 2 | 29.6 | 9.6 | 5.1 |
| day 3 | 31.1 | 11.1 | 5.9 |
| day 4 | 30.3 | 10.3 | 5.4 |
| day 5 | 26.6 | 6.6 | 3.5 |
| day 6 | 24.8 | 4.8 | 2.5 |
| day 7 | 24.6 | 4.6 | 2.4 |
| day 8 | 27.8 | 7.8 | 4.1 |
| day 9 | 26.8 | 6.8 | 3.6 |
| day 10 | 29.8 | 9.8 | 5.2 |
| IC on day 0 is from the added 17.5 mM Na$_2$CO$_3$. sample was taken before running the A501 belt. | | | |

**Table S4.** Total net CO$_2$ delivered by the *1g* system to the 12 L culture over 10 days

| Time, Days | Air/Belt-delivered CO$_2$ retained in 12 L of medium, g | CO$_2$ fixed by *Synechocystis* cells, g | Total CO$_2$ delivered by the 16 g of A501 (g) |
|---|---|---|---|
| 1 | 3.3 | | 3.3 |
| 2 | 5.1 | | 5.1 |
| 3 | 5.9 | | 5.9 |
| 4 | 5.4 | 1.1 | 6.5 |
| 5 | 3.5 | 2.6 | 6.1 |
| 6 | 2.5 | 4.1 | 6.6 |
| 7 | 2.4 | 5.4 | 7.9 |
| 8 | 4.1 | 6.1 | 10.2 |
| 9 | 3.6 | 5.3 | 8.9 |
| 10 | 5.2 | 4.3 | 9.5 |

*Synechocystis* cultivation was started on day 4.

**Figure S9.** Images of *Synechocystis* TE/Δ*slr1609* during cultivating with the *1g* system.

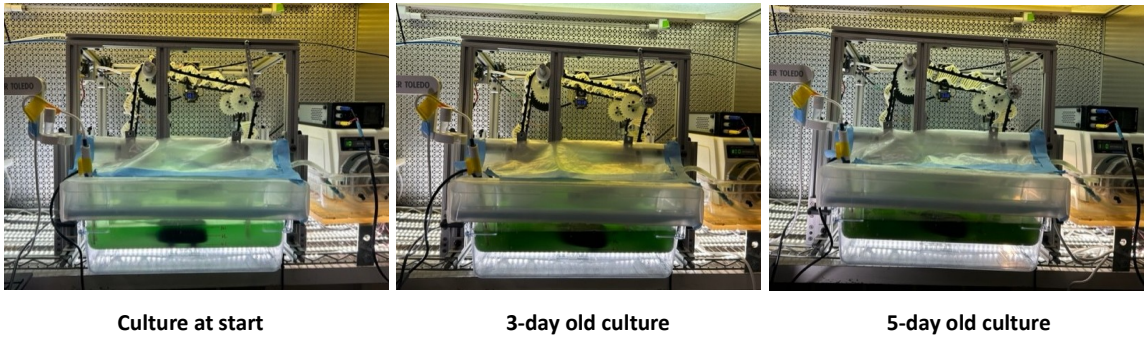

**Culture at start**  **3-day old culture**  **5-day old culture**

**Figure S9.** Images of the *Synechocystis* TE/Δ*slr1609* culture in the *1g* system during the first five days.

**Cultivation Trial #1 (6803)** During culture scale up, upon addition of $Na_2CO_3$ to raise the alkalinity and pH to ~10 in preparation for cultivation in the raceway pond significant biomass flocculation occurred where we lost a large portion of the biomass. After addition of 1 mL Antifoam A Concentrate (Sigma-Aldrich) per 100 L of culture, the culture recovered and was taken outside at the beginning of December to inoculate a 4.2 m$^2$ raceway pond to an optical density at 730 nm ($OD_{730}$) of ~0.3 filled to a 20-cm depth (~840 L) with the BG-11 background for trace elements and 5 mM sodium nitrate for N-source and alkalinity was adjusted to ~ 18 mM with the addition of $NaHCO_3$ and $Na_2CO_3$.

The seed culture was split equally between SPW18 (control) and SPW20 (*100g* system). No external $CO_2$ supplementation was provided to either pond, however, the control pond used a paddlewheel for circulation rather than a centrifugal pump like in the *100g* pond. Due to a miscommunication, only the *100g* pond was started at the target alkalinity. This was corrected at the reset on December 15, 2023 where we terminated SPW18, cleaned and sterilized the pond, and then split the culture from SPW 20 half and half between the two ponds

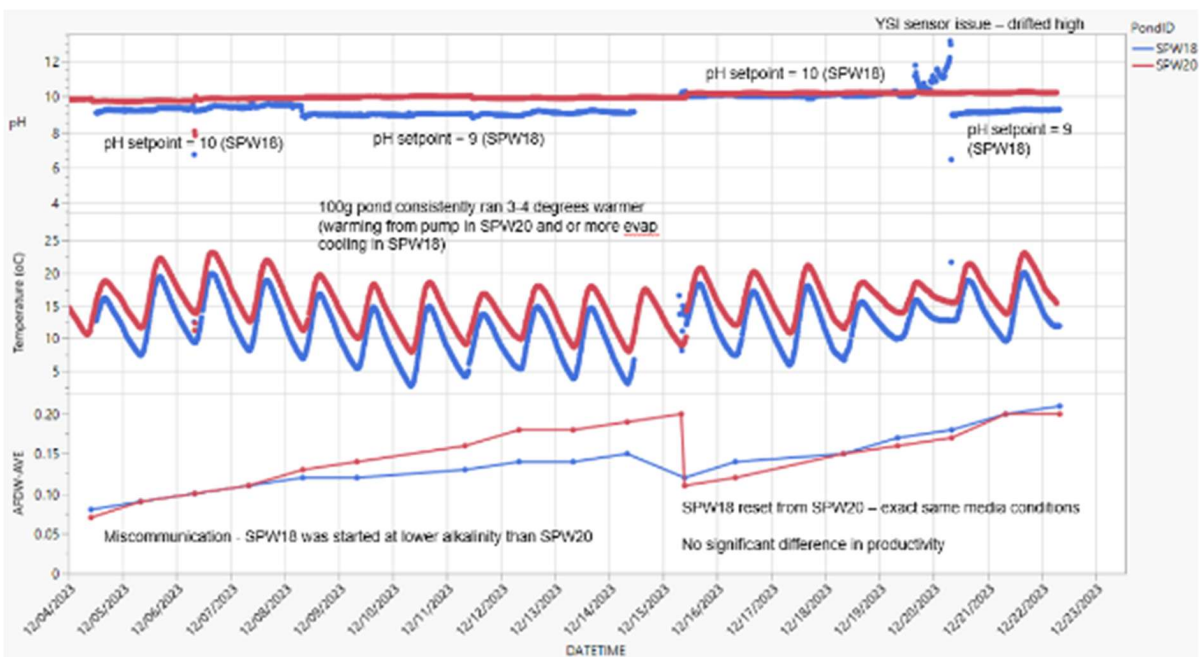

**Figure S10.** Cultivation trial #1 with *Synechocystis* 6803 in Dec. 2023 when grown with *100g* (SPW20) vs control (SPW18). Top graph: pH; middle graph: water temperature; and bottom graph AFDW (g/L). SPW20 was used to restart SPW18 on 12/15/2023.

The run began on 12/4/2023 and ran for 18 days until December 21, 2023. Although cultures in both ponds grew slowly during the first 10 days the *100g* pond grew somewhat higher (2.6 g/m$^2$-day) than the control (1.4 g/m$^2$-day) (**Figure S10**). The pH set point on SPW18 (control) was set originally to 10, but the pH of the medium always stayed below 10, so no $CO_2$ addition occurred. On December 8, the pH setpoint was dropped to 9 to give the control culture $CO_2$ and see if any improved growth occurred. None was observed and in fact, the culture became contaminated with amoeba and diatoms (**Figure S11**). Upon restart of SPW18 (control) at correct alkalinity with culture from SPW20 (*100g*), the second growout period from December 15 to 22 showed very similar growth rates for both ponds of 2.5–2.8 g/m$^2$-day. Interestingly, SPW20 (*100g*) pond never showed signs of contamination whereas the SPW18 (control) pond showed contamination on both growouts (**Figure S12**) in spite of identical medium pH and alkalinity during the second growout.

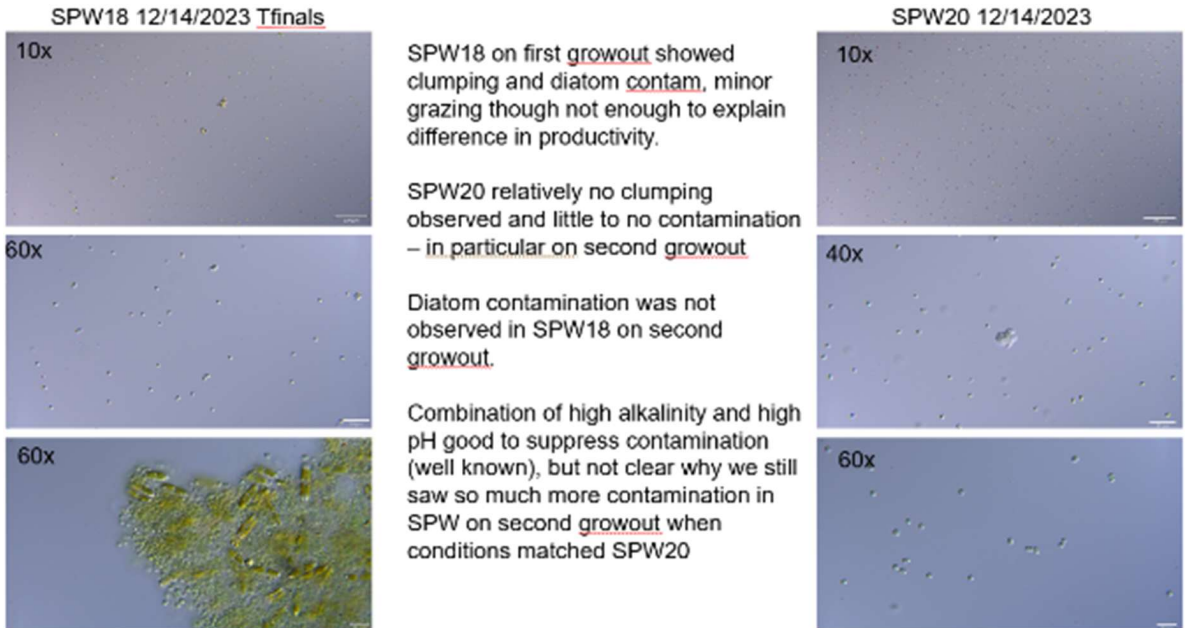

**Figure S11.** Optical microscopy comparisons for *Synechocystis* 6803 showing SPW18 (control) pond had contamination (left column) and SPW20 (*100g*) pond (right column) had little to no contamination present after 1 week of cultivation. SPW18 (control) pond was very clumpy with lots of diatom contamination in the clumps and also visible signs of amoeba grazing.

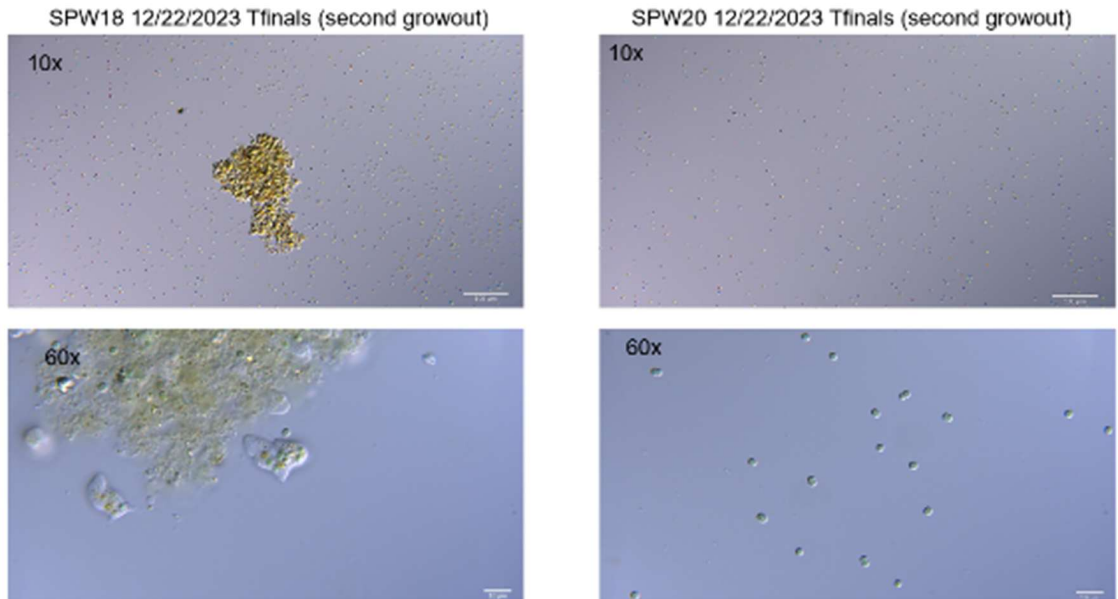

**Figure S12.** Optical microscopy comparisons for *Synechocystis* 6803 showing contamination in the control pond (SPW 18, left) whereas the *100g* pond (SPW20, right) had little to no contamination after 2 weeks total of cultivation on December 22. SPW18 control pond was very clumpy with lots of amoeba grazing, though no diatom contamination was observed.

**Cultivation Trial #2 (*Synechocystis* TE/Δ*slr1609* and *Desmodesmus*)**. On June 3, 2024 pond #28 was inoculated with *Synechocystis* TE/Δ*slr1609* into a BG-11 medium modified to reduce nitrate to 5 mM and adding 35 mM NaHCO$_3$ to ensure an excess of the moisture-swing-active anions HCO$_3^-$ and CO$_3^{2-}$ over other anions in the BG-11 medium required for growth (NO$_3^-$, PO$_4^{3-}$, SO$_4^{2-}$, Cl$^-$). As shown in **Figure S13** The culture did not look well after one day and, despite covering the culture with a shade cloth, it had died within a few days, most likely because of the low inoculation optical density (OD$_{750}$ ~0.1) that shocked the culture coming from columns; AzCATI is limited on the number of columns it can use for cultivating genetically modified strains. On June 13, a wild strain of the green algae *Desmodesmus* sp. (6–7 µm diameter) began growing, which had never popped up in a pond before at AzCATI under alkaline conditions. The strain had visible spikes/horns under the microscope (**Figure S13** inset), which indicates it was stressed, but it did grow rapidly from OD 0.1 to OD 0.3 under alkaline conditions. The pond was then split into another pond (pumped flow without a paddlewheel, as with *100g* pond) with CO$_2$ delivered by sparging and a pH setpoint of 9.5 where it grew up to OD 0.8, whereas the *100g* pond did not grow past OD 0.3 and the pH drifted up to 11, suggesting that growth in the *100g* pond was carbon limited.

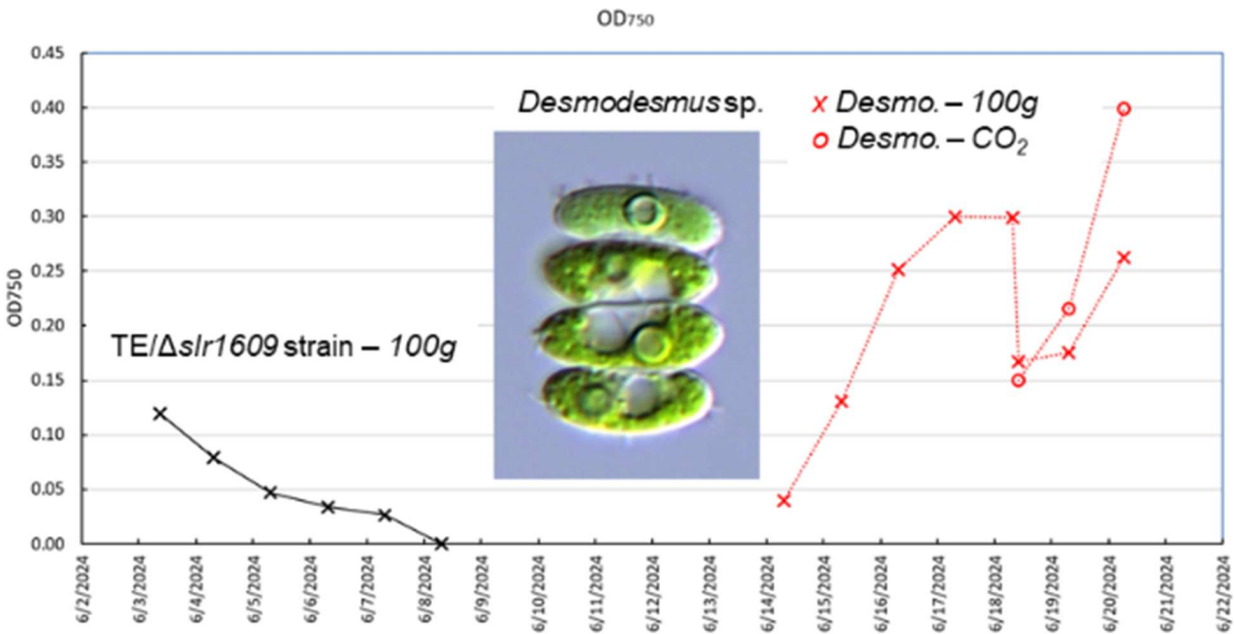

**Figure S13.** IC analysis results of the biotic media from the pond with the *100g* system from June 2 to 20, 2024.

**Figure S14** shows IC data from the *100g* pond, where initially *100g* is putting slightly more IC into the pond than the dying culture is removing. On June 9, since the culture had completely died, NaOH was added to raise the pH to 10 to continue testing *100g* CO$_2$ delivery into the abiotic BG-11 media. The IC delivery by *100g* then increased, which is because there was no longer carbon removal by the culture and because there is higher alkalinity to capture the delivered $CO_2$ into the medium. When *Desmodesmus* blew in and began growing on June 13 to June 21, IC levels declined indicating that during this time *Desmodesmus* consumed IC faster than the *100g* could deliver.

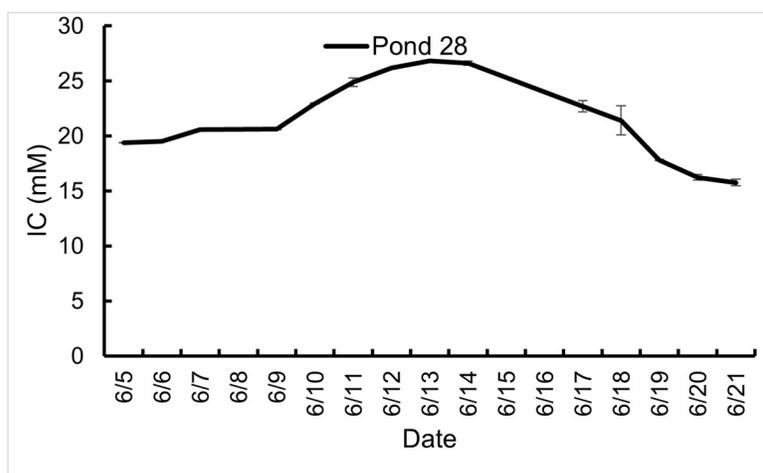

**Figure S14.** IC analysis results of the biotic media from the pond with the *100g* system.

**Cultivation Trial #3 (*Chlorella vulgaris* 1201).** Inorganic carbon (IC) levels in the pond with the *100g* system and the control pond from September 25 until the middle of December 2024 were measured and shown in **Figure S15**. Until early November, the ponds did not have microalgal or cyanobacterial cultures and was at about half-capacity in terms of the number of packets, thus expected to deliver ~50 g $CO_2$ per day. Initially, only one sorbent belt loop (A) was operated, which had 182 sorbent packets that were used in a previous cultivation trial over the summer that degraded their performance to about 50% of their initial capacity (after washing). On Oct. 18 through 20 the system lost power during a rainstorm, so was not operating. On October 21 an additional 118 new packets were installed on a second sorbent belt loop (B) to bring the total up to 300 packets. The rise in IC levels in the *100g* Pond 28 was much faster than in the control Pond 20, indicating that packets functioned well. The amount of $CO_2$ delivered during the abiotic trial agrees with what is expected based on the number of packets loaded and the performance of those packets (used vs new), releasing about 100 g $CO_2$ per day into the medium with all 300 packets.

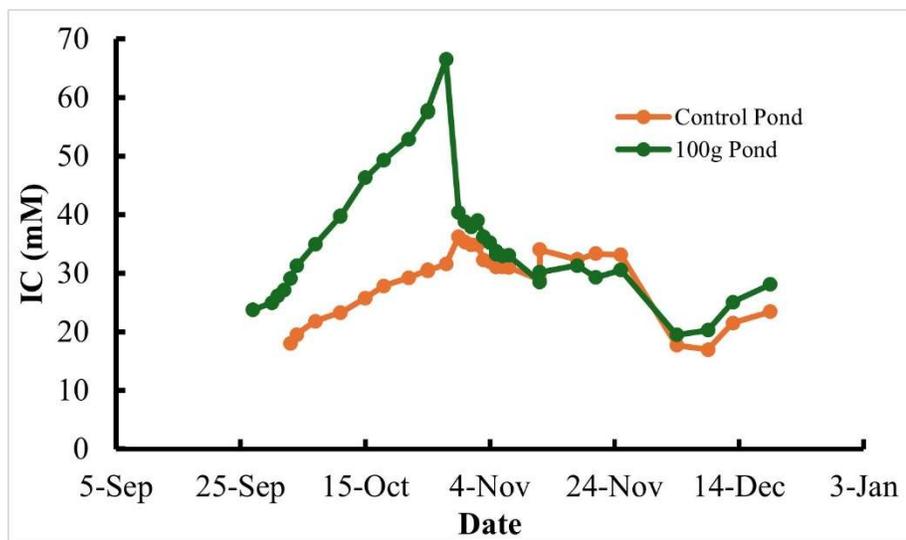

**Figure S15.** IC analysis results of the abiotic and biotic outdoor trials in fall 2024 with the *100g* system.

On October 30, the ponds were inoculated with *Chlorella vulgaris* 1201, which did not grow well in the *100g* pond. On Oct. 30, Nov. 6 and Nov. 18, the belt derailed overnight (so was not delivering $CO_2$) and was fixed the next day. On Nov. 6 the *100g* pond 28 was split into three ponds including the *100g* pond, the control pond, and a third pond to cultivate *Chlorella vulgaris* 1201 with external $CO_2$ supplementation. On Nov. 19 a derailment snapped the plastic chain, which took until Nov. 21 to repair. The system was stopped on Nov. 25 after the 1201 growth stagnated and the system remained off until Dec. 6 to clean the pond and the packets to try and restore the performance of the packets using 0.04% (w/v) bleach for 30 minutes, followed by an overnight soak in mild HCl (pH 3), followed in turn by soaking in 0.5 M $NaHCO_3$ for several days in a large 50-gallon drum. New medium was added to both ponds on Dec. 4th and the *100g* system was restarted on Dec. 6. Loop B derailed and was offline until Dec. 9. Loop A was repaired on Dec. 16 and 17. As indicated in **Figure S15**, little evidence of net IC accumulation was found during the *Chlorella vulgaris* 1201 cultivation trial, which we hypothesize was due to biofouling of the polymer by *Chlorella vulgaris* 1201.

**Cultivation Trial #4 (*Synechocystis* 6803).** The *100g* system containing two sorbent belt loops each made of 280 packets, which each contained two grams of A501 anion exchange resin, was installed on a 4.2 m² raceway pond (SPW-28) at AzCATI and operated with and without *Synechocystis* 6803 from Dec. 19, 2024 to Feb. 17, 2025 and compared to a control pond (SPW-20) without the AUDACity system; this pond only had passive $CO_2$ uptake from ambient air. Both ponds were circulated using a centrifugal pump rather than a paddlewheel. The *100g* and control ponds were started with a medium containing 17.5 mM $Na_2CO_3$ to evaluate abiotic $CO_2$ delivery of the *100g* system with 280 two-gram packets from Dec. 20 to January 13; NaOH was added to raise the pH to 10.25 whenever the pH of the medium dropped to 9.75. On January 16, both the control pond and the pond with the AUDACity system were inoculated with *Synechocystis* 6803 cultures in a modified BG-11 medium with $NaNO_3$ reduced to 5 mM and $NaHCO_3$ increased to 35 mM, and the system was run until February 17. $NaHCO_3$ was added to ensure a substantial excess relative to other nutrient anions in the medium that would otherwise reduce DAC performance. During abiotic and biotic operation, 50-mL liquid samples were collected daily to analyze inorganic carbon (IC) and organic carbon from the biomass and a selection of those were analyzed over that period. A Shimadzu Total Organic Carbon (TOC)-L analyzer was used to analyze IC by acidifying the sample and measuring the $CO_2$ released from the solution as well as by combusting the dry biomass.

As shown in **Figure S16**, the *100g* pond (SPW-28) had consistently higher IC levels than the control pond (SPW-20) that is attributed to the additional carbon being delivered by the sorbent and somewhat slower growth and IC removal by *Synechocystis* 6803. This is consistent with the pH data of the ponds in **Figure S17**: after inoculation the pH of the *100g* pond was consistently lower than that of the control pond, indicating that more $CO_2$ was delivered, especially later in the trial as the temperatures warmed up to increase the drying rate to allow for more $CO_2$ loading per cycle.

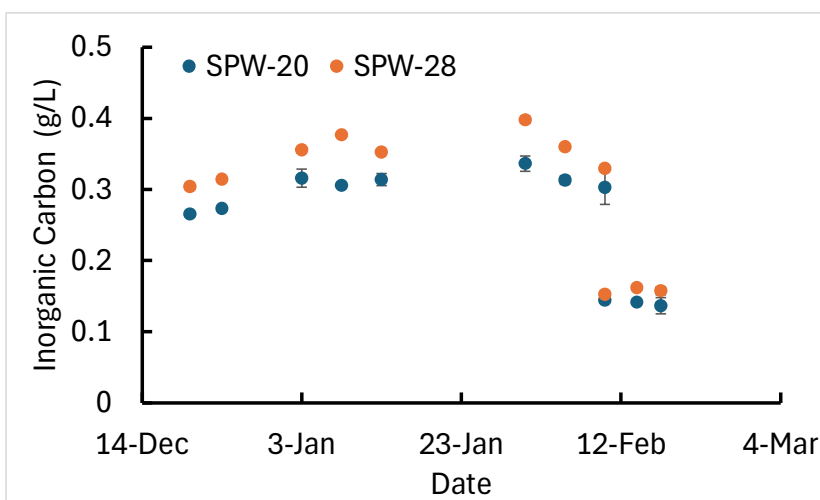

**Figure S16.** Inorganic carbon (IC) content in the *100g* pond (SPW-28) and the control pond (SPW-20) expressed as IC grams per liter of the liquid sample.

Some of the IC data in both ponds on December 24 and December 30 had significantly lower IC values than prior and subsequent readings and were omitted since they were not consistent with the pH data that showed a consistent $CO_2$ accumulation in both ponds over the period (**Figure S17**). The significant drop in IC on February 10 was when the biomass was harvested and fresh modified BG-11 medium with 35 mM $NaHCO_3$ was added to dilute the culture down to an $OD_{750}$ of 0.15. The pond was also harvested on January 24th (as shown in **Figure S18**), but unfortunately, no IC data were collected; thus, we hypothesize that the IC data between January 16 and January 31 underrepresent the amount of IC delivered to both ponds. The system was also offline from December 30 to January 2 and intermittently January 13–16 and 21–24 due to mechanical issues during operation. On January 14, 68 additional packets were installed and four damaged packets were removed to bring the total packet count up to 332 two-gram packets. On January 14, due to a significant mechanical failure, 52 damaged packets were removed reducing the total packet count to 280. **Figure S18** shows the organic carbon content from the analyzed biomass as calculated from the measured ash-free dry weight, where organic carbon makes up about 50% of AFDW.[45] These data show consistently higher organic carbon content in the control pond than the *100g* pond, although the difference appeared to close somewhat as the weather warmed up in mid-February. We hypothesize that the organic carbon content of the *100g* pond decreased immediately after inoculation due to sorbent fouling with biomass as the packets became visibly green.

However, additional factors may include stress from contacting the sorbent in addition to stresses from very low outdoor and pond temperatures (~5 °C).

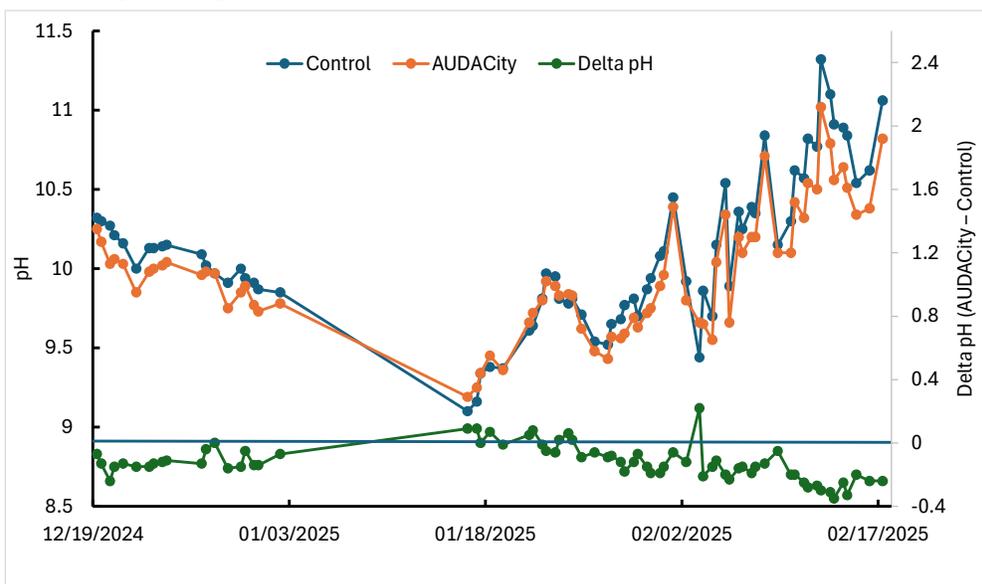

**Figure S17.** pH of the control pond (blue) and *100g* pond (orange) and the difference between the *100g* and control pond (green). Note: *100g* is also known as AUDACity.

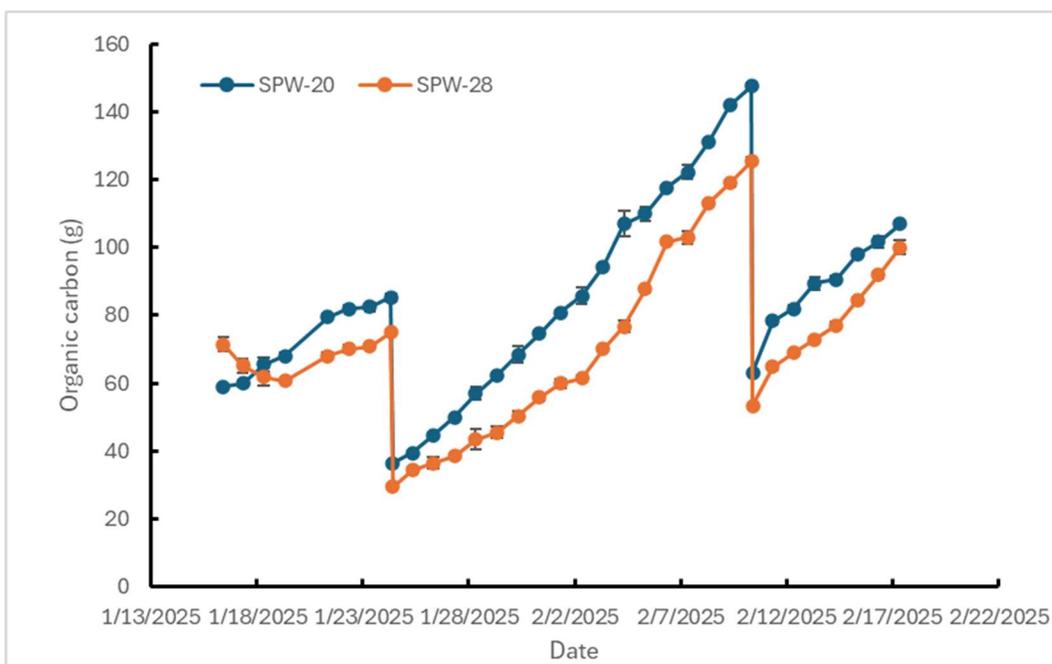

**Figure S18.** Organic carbon content calculated from the ash-free dry weight of the cyanobacterial biomass. The volume of culture ponds was 820–830 L. Organic carbon is expressed as total organic carbon (g) in the pond.

**Post-cultivation sorbent and mesh imaging.** Our hypothesis was that exopolysaccharides excreted by *Synechocystis* attached to the surface of the polymers to significantly reduce $CO_2$ transfer and ion exchange. Upon inspecting the polymers using scanning electron microscopy (SEM), the used polymer beads showed a much smoother surface compared to the new polymer beads (**Figure S19**). We hypothesize this is due to polysaccharide layers covering the surface of the beads.

Our attention next focused on the packets themselves: If polysaccharides may block the pores on the resin, might they also block the pores in the mesh packets that hold the A501 resin? As indicated in **Figure S20**, light microscopy showed that the pores of the mesh packets were covered with debris potentially resembling polysaccharides (**Figure 6**), while the new packets had clear pores (**Figure S20C**). Next the mesh packets were stained with PAS. As shown in **Figure S21**, new mesh packets did not stain with PAS, while the used packets developed an intense purple color, indicating the presence of polysaccharides. However, the used packets that had been washed in 1 mM HCl showed significantly less purple staining, indicating that a large part of the polysaccharides had been removed by HCl washing.

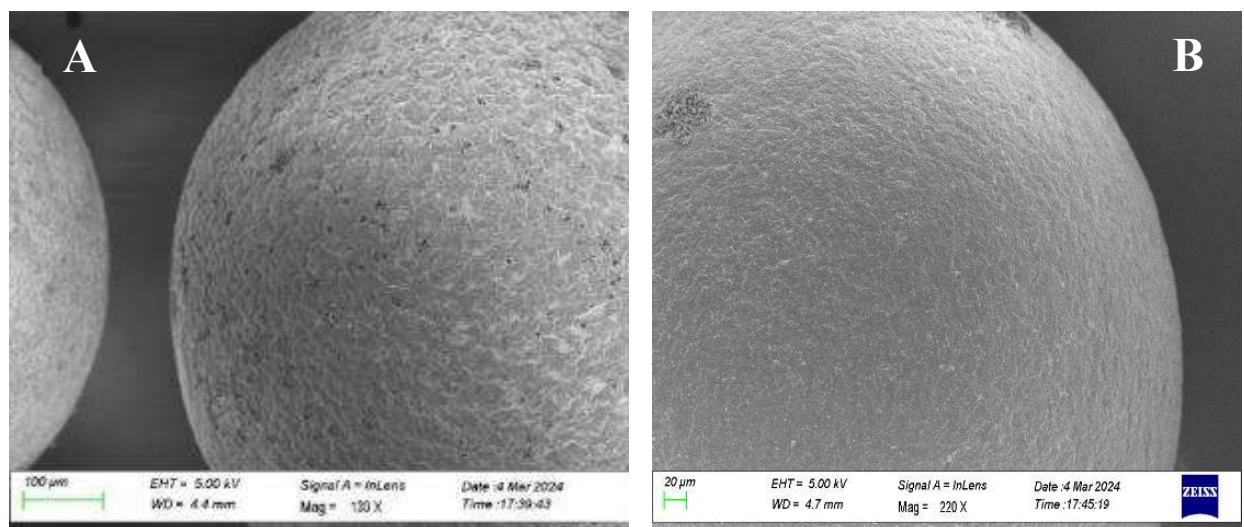

**Figure S19.** Scannin electron microscopy (SEM) images of new (A) and used (B) A501 polymer beads

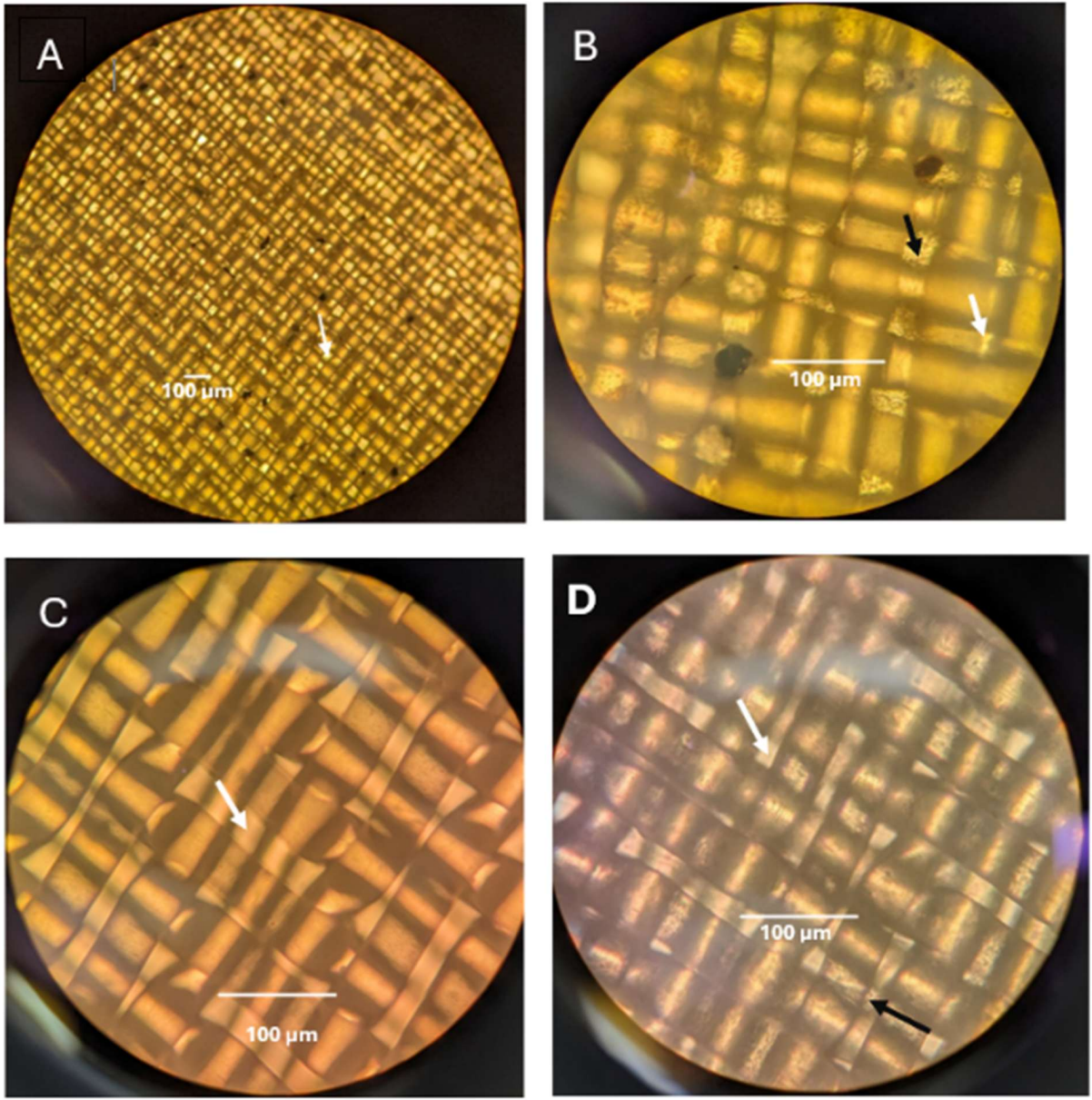

**Figure S20.** The mesh packets after use in the *100g* system for cultivation at AzCATI under a light stereoscope shown at 10X **(A)** and 40X **(B)** magnification **(C)** A 40X magnification of new mesh is shown for comparison. **(D)** A used mesh packet at 40X magnification after a HCl (pH 3) wash. White arrows show open pores; black arrows show obscured pores.

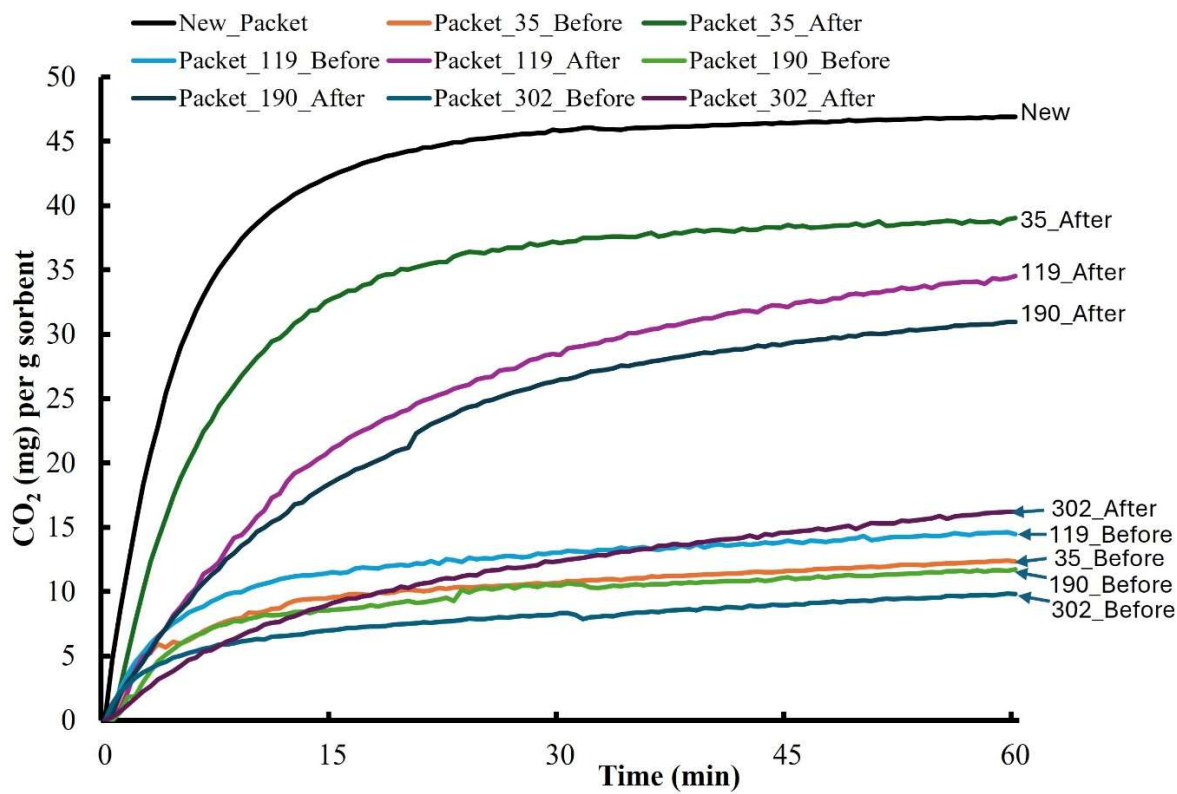

**Figure S21A.** $CO_2$ delivery into 10 mM $Na_2CO_3$ by sorbent packets used in outdoor trials for 0 (new), 35, 119, 190 or 302 days before and after a cleaning protocol.

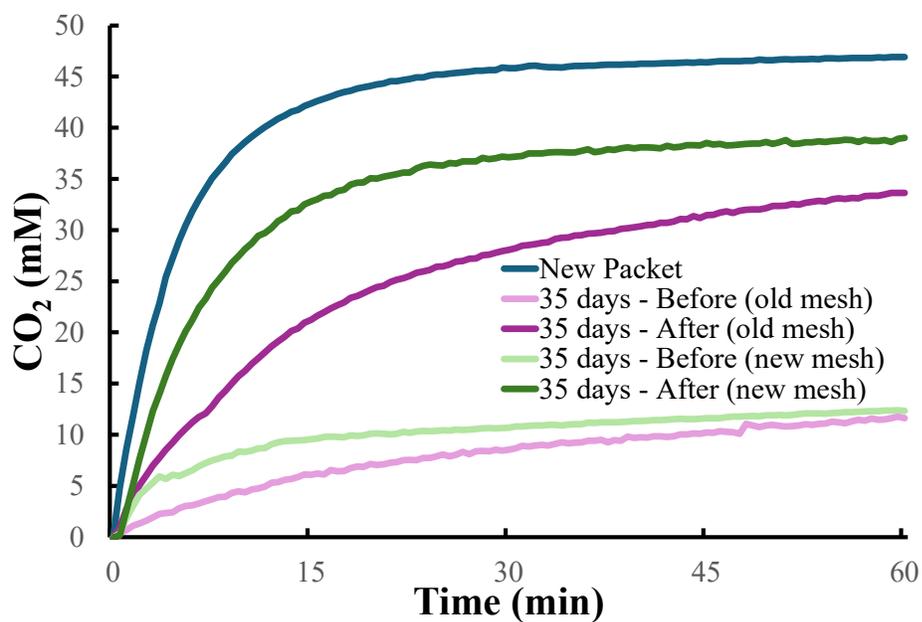

**Figure S21B.** $CO_2$ delivery into 10 mM $Na_2CO_3$ by sorbent packets used in outdoor trials for 35 days in their original mesh packet and when the sorbent was transferred to a new mesh packet, before and after a cleaning protocol.

**Mesh packet damage analysis.** Packets were examined individually for damage and sorbent loss. **Figure S22** shows a packet with damage locations specified. Damage location was recorded as (1) damage along the interior seam of each tube holding the beads, either along the parallel seams (horizontal seal lines) or ends (vertical seal lines) causing sorbent loss; (2) damage to the "attachment points" specifically where packets attached to the roller chain belt and did not cause sorbent loss; or (3) damage to the "tab" where two adjacent packets were snapped together to form a double-wide packet (spanning the parallel roller chains), usually at a sacrificial outer seam such that the inner seam would prevent sorbent loss. We hypothesize that damage to packets at the tab or attachment points was caused by mechanical forces from the motor driving only one chain and as such was designated as stress damage. Damage to packets along the parallel or end seams are hypothesized to be due to deterioration and weakening of the plastic materials at their weakest point, the thermally sealed seams, and as such was designated as degradation damage.

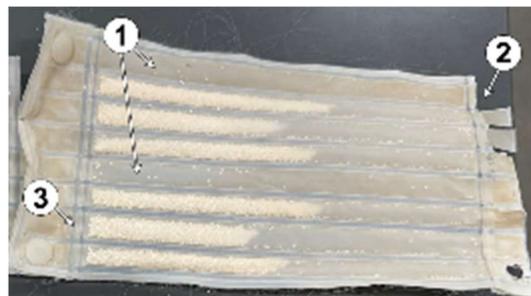

**Figure S22: Damaged Packet.** (1) Empty tubes with no obvious stress damage contain small rips along parallel seams (horizontal seal lines) or ends (vertical). (2) Attachment point damage. (3) Location where tab damage would occur.

Table S5. Tabulation of sorbent packet usage duration and exposure to cultivation, wet/dry cycles and UV.

| Install Date | Jan. 1, 2025 | Oct 21, 2024 | July 18, 2024 | June 21, 2024 | May 7, 2024 | Mar. 28, 2024 |
|---|---|---|---|---|---|---|
| # of Active Days | 35 | 119 | 190 | 217 | 262 | 302 |
| Cumulative Wet/Dry Cycles | 142.5 ± 3.5 | 411.5 ± 3.5 | 708.5 ± 8.5 | 939.5 ± 8.5 | 1253.5 ± 8.5 | 1379.5 ± 8.5 |
| # of Days in Cultivation | 33 | 60 | 60 | 68 | 78 | 78 |
| Cumulative UV Exposure (MJ m$^{-2}$) | 0.027 ± 0.007 | 0.19 ± 0.05 | 0.8 ± 0.2 | 1.0 ± 0.25 | 1.5 ± 0.36 | 1.8 ± 0.4 |

**Figure S23** shows the FTIR spectra (**A**) and TGA (**B**) of sorbent used in the *100g* system for 0, 190 and 302 days.

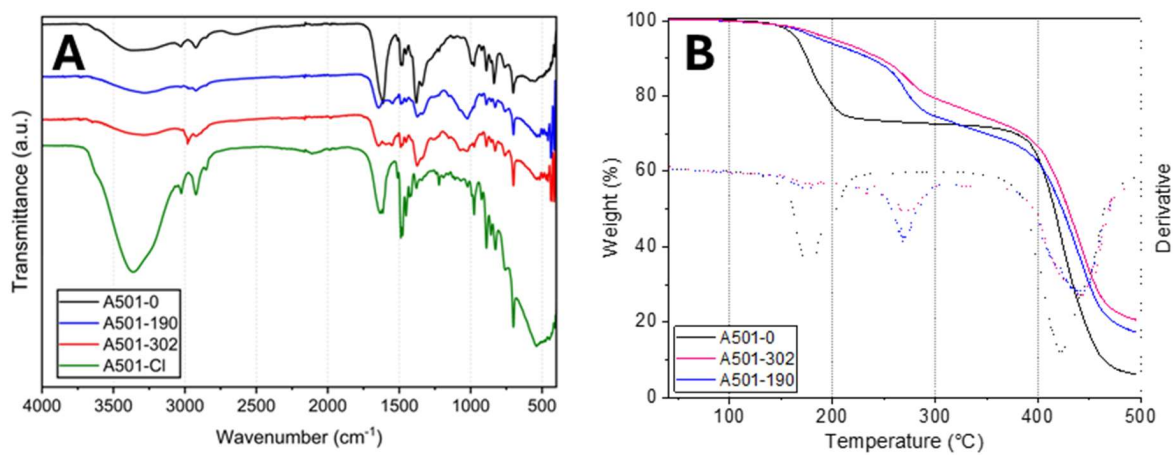

**Figure S23.** FTIR spectra (**A**) and TGA (**B**) of sorbent used in the *100g* system for 0, 190 and 302 days. FTIR is also shown for an unused sorbent as it comes from the manufacturer, with Cl$^-$ counterion that is inactive for $CO_2$ capture.

**Scanning electron microscopy (SEM) analysis.** Sorbent beads used for 0, 190 and 302 days were imaged by SEM to investigate changes to the sorbents microporosity and surface (**Figure S24**).

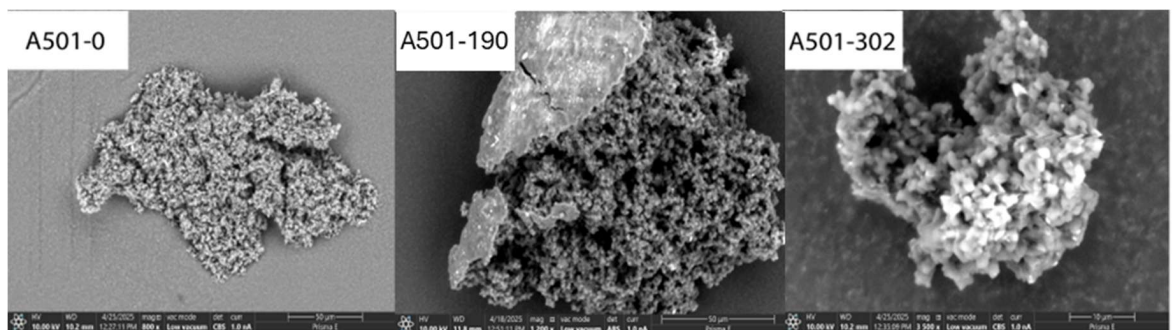

**Figure S24.** SEM micrographs showing the microporous structure of A501 sorbent beads used for 0, 190 and 302 days.

**Technoeconomic and life cycle analysis (TEA/LCA) calculations and assumptions.**

This section documents key assumptions used in the integrated biorefinery process model, building on those from the DAC subprocess model with assumptions for the HTL model (**Table S6**), pond model (**Table S7**) and seasonal inputs pond model (**Table S8**).

**Table S6**: Inputs to Chen et al. HTL model

| Input | Value | Units |
|---|---|---|
| Lipids | 25.00% | Percent of AFDW |
| Carbohydrates | 35.00% | Percent of AFDW |
| Proteins | 40.00% | Percent of AFDW |
| Ash | 5% | Percent of dry weight |
| Algal Productivity | | g AFDW m$^{-2}$ d$^{-1}$ |
| Facility Size | 5000 | Wetted acres |

AFDW: Ash free dry weight

**Table S7**: Inputs to Davis et al. Pond model

| | | |
|---|---|---|
| Carbon | 52% | Percent of AFDW |
| Hydrogen | 7.84% | Percent of AFDW |
| Oxygen | 28.64% | Percent of AFDW |
| Nitrogen | 6.08% | Percent of AFDW |
| Sulphur | 0.20% | Percent of AFDW |
| Phosphorous | 0.22% | Percent of AFDW |
| Ash Content | 5.00% | Percent of dry weight |
| Facility size (based on cultivation area) | 5000 | acres |
| Module size (based on cultivation area) | 100 | acres |
| Individual pond size | 10-acre | |
| Pond liner coverage | Minimally Lined | |
| Pond depth | 20 | cm |
| Pond harvest concentration | 0.5 | g/L |
| Fresh or saline water cultivation | Freshwater | |
| Total pond volume | 4,046,860 | m^3 |
| $CO_2$ uptake efficiency in ponds | 90% | |
| $CO_2$ uptake efficiency in PBRs | 99% | |
| Excess $NH_3$ to ponds | 20% | |

| | | | |
|---|---|---|---|
| Excess DAP to ponds | | 10% | |
| Production ponds initial concentration at inoculation | | 0.10 | g/L |
| Outlet concentration of 1st-stage inoculum (tubular PBR) | | 1.5 | g/L |
| Outlet concentration of 2nd-stage inoculum (closed ponds) | | 0.5 | g/L |
| Outlet concentration of 3rd-stage inoculum (lined open ponds) | | 0.5 | g/L |
| Pond up-time between re-inoculation | | 20 | days |
| Primary dewatering outlet concentration | | 10 | g/L |
| Primary dewatering harvest efficiency | | 90% | |
| Residence time for settling tank (primary dewatering) | | 4 | hr |
| Secondary dewatering outlet concentration | | 130 | g/L |
| Secondary dewatering harvest efficiency | | 99.5% | |
| Tertiary dewatering outlet concentration | | 200 | g/L |
| Tertiary dewatering harvest efficiency | | 97% | |
| Biomass loss during short-term product storage | | 1% | |

**Table S8**: Seasonal inputs to Davis et al. Pond model

| | SUMMER (JUN, JUL, AUG) | FALL (SEP, OCT, NOV) | WINTER (DEC, JAN, FEB) | SPRING (MAR, APRL, MAY) | Yearly Average | Unit |
|---|---|---|---|---|---|---|
| Productivity | 35.0 | 24.9 | 11.7 | 28.5 | 25.0 | g/m²/day |
| Evaporation | 0.09 | 0.035 | 0.035 | 0.189 | 0.1 | cm/day |
| Blowdown For Facility | 304135 | 116817 | 111986 | 514894 | 261958.0 | kg/hr |